\documentclass[a4paper, reqno, 12pt]{amsart}

\usepackage[left=2.5cm, right=2.5cm, top=3cm, bottom=3cm]{geometry}

\usepackage{amsaddr}
\usepackage{amssymb}
\usepackage{amsfonts}
\usepackage{mathrsfs}
\usepackage[T1]{fontenc}
\usepackage[USenglish]{babel}
\usepackage{varioref}
\usepackage{multirow}
\usepackage{array,multicol}
\usepackage{bigdelim}
\usepackage{bigstrut}
\usepackage{bm}
\usepackage{units}
\usepackage{bbold}
\usepackage[font=small, labelfont=bf]{caption}
\usepackage[list=true, labelformat=brace, font=small, labelfont=bf]{subcaption}
\usepackage{graphicx}
\usepackage{csquotes}

\usepackage[breaklinks=true
]{hyperref}
\hypersetup{pdftitle={F.Nill: Symmetries and normalization in 3-compartment epidemic models. I: The replacement number dynamics.}}

\usepackage[backend=biber,style=authoryear-comp, dashed=false]{biblatex}

\usepackage{enumitem}
\setlist[itemize]{align=parleft,left=0pt..1.5em}
\setlist{noitemsep}

\usepackage{url}

\def\customauthor{\empty}
\def\customdate{\empty}
\let\oldauthor\author
\renewcommand{\author}[1]{\def\customauthor{#1}}
\renewcommand{\date}[1]{\def\customdate{#1}}
\oldauthor{\customauthor\\\customdate}

\theoremstyle{definition}
\newtheorem{definition}{Definition}[section]

\theoremstyle{plain}
\newtheorem{theorem}[definition]{Theorem}
\newtheorem{lemma}[definition]{Lemma}
\newtheorem{corollary}[definition]{Corollary}
\newtheorem{proposition}[definition]{Proposition}

\theoremstyle{remark}
\newtheorem{remark}[definition]{Remark}

\numberwithin{equation}{section}

\newcommand{\ncm}{\newcommand}
\ncm{\rncm}{\renewcommand}

\ncm{\lb}[1]{\label{#1}}

\rncm{\sec}{\setc{0}\section}
\ncm{\bsn}{\bigskip\noindent}
\ncm{\msn}{\medskip\noindent}
\ncm{\ssn}{\smallskip\noindent}
\ncm{\beq}{\begin{equation}}
\ncm{\beqnon}{\begin{equation*}}
\ncm{\eeq}{\end{equation}}
\ncm{\eeqnon}{\end{equation*}}
\ncm{\bea}{\begin{eqnarray}}
\ncm{\beanon}{\begin{eqnarray*}}
\ncm{\eea}{\end{eqnarray}}
\ncm{\eeanon}{\end{eqnarray*}}

\ncm{\ba}{\begin{array}}
\ncm{\ea}{\end{array}}

\ncm{\fns}{\footnotesize}

\DeclareMathOperator\sign{sign}
\DeclareMathOperator\const{const.}

\DeclareMathOperator\diag{diag}

\DeclareMathOperator\age{age}

\ncm{\scenA}{\ensuremath{\mathrm{(A)}}}
\ncm{\scenB}{\ensuremath{\mathrm{(B)}}}
\ncm{\scenC}{\ensuremath{\mathrm{(C)}}}
\ncm{\scenD}{\ensuremath{\mathrm{(D)}}}
\ncm{\scenE}{\ensuremath{\mathrm{(E)}}}

\ncm{\scenAp}{\ensuremath{\mathrm{(A^+)}}}
\ncm{\scenBp}{\ensuremath{\mathrm{(B^+)}}}
\ncm{\scenCp}{\ensuremath{\mathrm{(C^+)}}}
\ncm{\scenDp}{\ensuremath{\mathrm{(D^+)}}}
\ncm{\scenEp}{\ensuremath{\mathrm{(E^+)}}}

\ncm{\scenAm}{\ensuremath{\mathrm{(A^-)}}}
\ncm{\scenBm}{\ensuremath{\mathrm{(B^-)}}}
\ncm{\scenCm}{\ensuremath{\mathrm{(C^-)}}}
\ncm{\scenDm}{\ensuremath{\mathrm{(D^-)}}}

\ncm{\scenXpm}{\ensuremath{\mathrm{(X_\pm)}}}

\newcommand{\RR}{\ensuremath{\mathbb{R}}}
\newcommand{\RRN}{\ensuremath{\mathbb{R}_{\geq 0}}}
\ncm{\NN}{\ensuremath{\mathbb{N}}}
\ncm{\ZZ}{\ensuremath{\mathbb{Z}}}
\ncm{\GG}{\ensuremath{\mathbb{G}}}
\rncm{\SS}{\ensuremath{\mathbb{S}}}
\ncm{\bone}{\ensuremath{\mathbb{1}}}
\newcommand{\II}{{ \mathbb{I}}}
\newcommand{\bA}{ \mathbf{A}}
\newcommand{\bB}{ \mathbf{B}}
\newcommand{\bC}{ \mathbf{C}}
\newcommand{\bD}{ \mathbf{D}}
\newcommand{\bE}{ \mathbf{E}}
\newcommand{\bF}{ \mathbf{F}}
\newcommand{\bL}{ \mathbf{L}}

\newcommand{\bj}{ \mathbf{j}}

\newcommand{\bY}{ \mathbf{Y}}

\newcommand{\bS}{ \mathbf{S}}
\newcommand{\bT}{ \mathbf{T}}
\newcommand{\bV}{ \mathbf{V}}
\newcommand{\bM}{ \mathbf{M}}

\newcommand{\bp}{ \mathbf{p}}

\newcommand{\bg}{ \mathbf{g}}

\newcommand{\bx}{ \mathbf{x}}
\newcommand{\by}{ \mathbf{y}}

\newcommand{\bal}{ \bm{\alpha}}
\newcommand{\bbe}{ \bm{\beta}}
\newcommand{\bga}{ \bm{\gamma}}
\newcommand{\bth}{ \bm{\theta}}
\newcommand{\bvth}{ \bm{\vartheta}}

\newcommand{\bLa}{ \bm{\Lambda}}
\newcommand{\bGa}{ \bm{\Gamma}}

\newcommand{\bnu}{ \bm{\nu}}

\ncm{\pt}{\bp_\tau}
\newcommand{\bfa}{ \mathbf{a}}
\newcommand{\bfe}{ \mathbf{e}}

\newcommand{\bff}{\mathbf{f}}

\newcommand{\bzero}{\mathbf{0}}

\newcommand{\A}{\ensuremath{{\mathcal A}}}

\newcommand{\B}{\ensuremath{{\mathcal B}}}

\newcommand{\C}{\ensuremath{{\mathcal C}}}
\newcommand{\D}{\ensuremath{{\mathcal D}}}
\newcommand{\E}{\ensuremath{{\mathcal E}}}

\newcommand{\G}{\ensuremath{{\mathcal G}}}
\newcommand{\K}{\ensuremath{{\mathcal K}}}

\newcommand{\T}{\ensuremath{{\mathcal T}}}

\newcommand{\Z}{\ensuremath{{\mathcal Z}}}
\rncm{\P}{\ensuremath{{\mathcal P}}}
\ncm{\PPe}{\P_\epsilon}

\rncm{\S}{\ensuremath{{\mathcal S}}}

\newcommand{\R}{\ensuremath{{\mathcal R}}}

\renewcommand{\L}{\ensuremath{{\mathcal L}}}
\ncm{\Labc}{\L_{a,b,c}}
\newcommand{\I}{\ensuremath{{\mathcal I}}}

\newcommand{\M}{\ensuremath{{\mathcal M}}}

\rncm{\O}{\mathcal{O}}

\ncm{\Me}{\M_\epsilon}

\ncm{\Pph}{\P_{\mathrm{phys}}}
\ncm{\Tph}{\T_{\mathrm{phys}}}
\ncm{\bTph}{\bar{\T}_{\mathrm{phys}}}
\ncm{\Iend}{\I_{\mathrm{end}}}
\ncm{\Ta}{\T_{\bm{a}}}
\ncm{\Aph}{\A_{\mathrm{phys}}}
\ncm{\Abio}{\A_{\mathrm{bio}}}
\ncm{\Asig}{\A_{\mathrm{split}}}
\ncm{\Csig}{\C_{\mathrm{split}}}
\ncm{\Dsig}{\D_{\mathrm{split}}}
\ncm{\Aone}{\A_{\mathrm{Model-1}}}
\ncm{\Atwo}{\A_{\mathrm{Model-2}}}
\ncm{\Asirs}{\A_{\mathrm{SIRS}}}
\ncm{\Asiso}{\A_{\mathrm{SIS-1}}}
\ncm{\Asist}{\A_{\mathrm{SIS-2}}}
\ncm{\Asisp}{\A_{\mathrm{SIS+}}}
\ncm{\Asism}{\A_{\mathrm{SIS-}}}
\ncm{\Asispm}{\A_{\mathrm{SIS\pm}}}
\ncm{\Asisj}{\A_{\mathrm{SIS-j}}}
\ncm{\Aheth}{\A_{\mathrm{Heth}}}
\ncm{\Dheth}{\D_{\mathrm{Heth}}}
\ncm{\Kheth}{\K_{\mathrm{Heth}}}
\ncm{\Cph}{\C_{\mathrm{phys}}}
\ncm{\Cbio}{\C_{\mathrm{bio}}}
\ncm{\Cone}{\C_{\mathrm{Model-1}}}
\ncm{\Ctwo}{\C_{\mathrm{Model-2}}}
\ncm{\Csirs}{\C_{\mathrm{SIRS}}}
\ncm{\Dph}{\D_{\mathrm{phys}}}
\ncm{\bDph}{\bar{D}_{\mathrm{phys}}}
\ncm{\DBA}{\D_{AB}}
\ncm{\DAB}{\D_{AB}}
\ncm{\Dbio}{\D_{\mathrm{bio}}}
\ncm{\Dbionu}{\D_{{\mathrm{bio},\nu}}}
\ncm{\Dbioz}{\D_{{\mathrm{bio},0}}}
\ncm{\Done}{\D_{\mathrm{Model-1}}}
\ncm{\Dtwo}{\D_{\mathrm{Model-2}}}
\ncm{\DII}{\D_{\RN{2}}}
\ncm{\Dsirs}{\D_{\mathrm{SIRS}}}
\ncm{\Ksirs}{\K_{\mathrm{SIRS}}}
\ncm{\Kbio}{\K_{\mathrm{bio}}}
\ncm{\Kbionu}{\K_{{\mathrm{bio},\nu}}}
\ncm{\Lbio}{\L_{\mathrm{bio}}}
\ncm{\Gdil}{G_{\mathrm{dil}}}
\ncm{\GX}{G_{X}}
\ncm{\GI}{G_{I}}

\ncm{\tAph}{\tilde{\A}_{\mathrm{phys}}}
\ncm{\tAbio}{\tilde{\A}_{\mathrm{bio}}}
\ncm{\tAone}{\tilde{\A}_{\mathrm{Model-1}}}
\ncm{\tAtwo}{\tilde{\A}_{\mathrm{Model-2}}}
\ncm{\tAsirs}{\tilde{\A}_{\mathrm{SIRS}}}
\ncm{\TABC}{T(\bA,\bB,\bC)}
\ncm{\Tabc}{\TABC^{\leq 1}}

\ncm{\yph}{y_{\mathrm{phys}}}
\ncm{\Padm}{P_{\mathrm{adm}}}
\ncm{\Plow}{P_{\mathrm{low}}}
\ncm{\Phigh}{P_{\mathrm{high}}}
\ncm{\plow}{p_{\mathrm{low}}}
\ncm{\phigh}{p_{\mathrm{high}}}
\ncm{\Pc}{\P_{\mathrm{cut}}}
\ncm{\Reff}{X_{\mathrm{rep}}}
\ncm{\Reffo}{X_{\mathrm{rep,0}}^*}
\ncm{\Reffe}{X_{\mathrm{rep,end}}^*}
\ncm{\Ie}{I_{\mathrm{end}}^*}
\ncm{\Se}{I_{\mathrm{end}}^*}
\ncm{\dReff}{\dot{X}_{\mathrm{rep}}}
\ncm{\Tosc}{T_{\mathrm{osc}}}
\ncm{\Timm}{T_{\mathrm{imm}}}
\ncm{\Tinf}{T_{\mathrm{inf}}}
\ncm{\Thalf}{T_{\mathrm{half}}}

\ncm{\rvac}{r_{\mathrm{vac}}}
\newcommand{\al}{\alpha}
\newcommand{\la}{\lambda}
\newcommand{\be}{\beta}

\newcommand{\ga}{\gamma}

\newcommand{\ep}{\epsilon}
\newcommand{\vep}{\varepsilon}
\newcommand{\om}{\omega}

\newcommand{\ka}{\kappa}
\renewcommand{\th}{\theta}
\ncm{\OP}{\Omega_{\PP,\ep}}
\ncm{\oG}{\omega_\G}
\ncm{\ome}{\omega_\ep}
\ncm{\phit}{\varphi_\tau}
\ncm{\p}{\psi}

\ncm{\Aal}{A_\alpha}
\ncm{\Bal}{B_\alpha}
\ncm{\sal}{\sigma_\alpha}

\rncm{\k}{\kappa}
\ncm{\an}{a_\nu}
\newcommand{\re}{{\,\hbox{$\textstyle\triangleright$}\,}}

\newcommand{\li}{{\,\hbox{$\textstyle\triangleleft$}\,}}

\newcommand{\id}{{\rm id}}

\newcommand{\one}{{\mathbf 1}}

\newcommand{\bra}{\langle}
\newcommand{\ket}{\rangle}
\def\End{\mbox{End}\,}

\def\Hom{\mbox{Hom}\,}

\def\cros{\,\raise1.9pt\hbox{$\scriptscriptstyle  > $}\!
          \raise1.5pt\hbox{$\scriptstyle\triangleleft$}\,}
\def\>cros{\cros}
\def\<cros{\,\raise1.5pt\hbox{$\scriptstyle\triangleright$}\!
           \raise1.9pt\hbox{$\scriptscriptstyle < $}\,}
\ncm{\veq}{{\scriptstyle\Vert}}

\ncm{\dR}{\partial_R}
\ncm{\dS}{\partial_S}
\ncm{\dI}{\partial_I}
\ncm{\dD}{\partial_D}
\ncm{\dN}{\partial_N}
\ncm{\dM}{\partial_M}
\ncm{\dX}{\partial_X}
\ncm{\dq}{\partial_q}
\ncm{\dx}{\partial_x}
\ncm{\dy}{\partial_y}
\ncm{\parH}{\partial H}
\ncm{\parHe}{\partial H_{\epsilon}}
\ncm{\parq}{\partial q}
\ncm{\parp}{\partial p}

\ncm{\rto}{\rightarrow}
\ncm{\mto}{\longmapsto}
\ncm{\lto}{\longrightarrow}
\ncm{\Lto}{\Longrightarrow}
\ncm{\lra}{\leftrightarrow}
\ncm{\LRA}{\Leftrightarrow}
\ncm{\LLRA}{\Longleftrightarrow}
\ncm{\LRa}{\Leftrightarrow}
\ncm{\LLRa}{\Longleftrightarrow}

\ncm{\tOP}{\tilde{\Omega}_{\PP,\ep}}
\ncm{\toe}{\tilde{\omega}_\ep}
\ncm{\tHe}{\tilde{H}_{\epsilon}}
\ncm{\tH}{\tilde{H}}
\ncm{\tV}{\tilde{V}}
\ncm{\tK}{\tilde{K}}
\ncm{\tE}{\tilde{E}}
\ncm{\tA}{\tilde{\A}}
\ncm{\tP}{\tilde{\P}}
\ncm{\tbF}{\tilde{\bF}}
\ncm{\rt}{\tilde{r}_0}
\ncm{\tr}{\tilde{r}_0}
\ncm{\tga}{\tilde{\gamma}}
\ncm{\tal}{\tilde{\alpha}}
\ncm{\tGa}{\tilde{\Gamma}}
\ncm{\Nt}{\tilde{N}}
\ncm{\tHPe}{\tilde{H}_{\PP,\epsilon}}
\ncm{\tre}{\tilde{\rho}_{\epsilon}}
\ncm{\tq}{\tilde{q}}
\ncm{\tp}{\tilde{p}}
\ncm{\ta}{\tilde{a}}
\ncm{\tb}{\tilde{b}}
\ncm{\tbe}{\tilde{\beta}}
\ncm{\tc}{\tilde{c}}
\ncm{\td}{\tilde{d}}
\ncm{\tep}{\tilde{\vep}}
\ncm{\etq}{e^{\tilde{q}}}
\ncm{\etp}{e^{\tilde{p}}}
\ncm{\ttau}{\tilde{\tau}}
\ncm{\hH}{\hat{H}}
\ncm{\hV}{\hat{V}}
\ncm{\hK}{\hat{K}}
\ncm{\hKs}{\hat{K}_\sigma}
\ncm{\hHs}{\hat{H}_\sigma}
\ncm{\hvc}{\hat{v}_C}
\ncm{\hP}{\hat{\P}}
\ncm{\hA}{\hat{\A}}
\ncm{\hAph}{\hat{\A}_{\mathrm{phys}}}
\ncm{\hB}{\hat{\B}}
\ncm{\hC}{\hat{\C}}
\ncm{\hCph}{\hat{\C}_{\mathrm{phys}}}
\ncm{\hD}{\hat{\D}}
\ncm{\hphi}{\hat{\phi}}
\ncm{\crho}{\check{\rho}}
\ncm{\Reflat}{R_1^{\ \flat}}

\ncm{\ok}{\checkmark}
\ncm{\s}{\mathsf{s}}
\ncm{\g}{\mathsf{g}}
\ncm{\h}{\mathsf{h}}
\ncm{\HIG}{H_\alpha}
\ncm{\Hal}{H_\alpha}
\ncm{\oIG}{\omega_\alpha}
\ncm{\HPLV}{H_{\mathrm {pLV}}}
\ncm{\omPLV}{\om_{\mathrm {pLV}}}
\ncm{\Hreg}{H^{\mathrm {reg}}}
\ncm{\omreg}{\om^{\mathrm {reg}}}
\ncm{\HPLVreg}{H_{\mathrm {pLV}}^{\mathrm {reg}}}
\ncm{\omPLVreg}{\omega_{\mathrm {pLV}}^{\mathrm{reg}}}
\ncm{\Halreg}{H_{\alpha}^{\mathrm {reg}}}
\ncm{\omalreg}{\omega_{\alpha}^{\mathrm {reg}}}
\ncm{\xnull}{x_{\mathrm {null}}}
\ncm{\Lt}{L_\tau}
\ncm{\Lmin}{L_{\min}}
\ncm{\dLt}{\dot{L}_\tau}
\ncm{\rv}{a_\mathrm{vac}}

\newcommand{\Del}{\Delta}
\ncm{\DBo}{\D_{\Del B>0}}
\ncm{\DelB}{\Del B}

\newcommand{\Eqref}[1]{Eq. \eqref{#1}}

\ncm{\ulim}[1]{\underset{#1}{\lim}}
\ncm{\secref}[1]{Section \ref{#1}}
\ncm{\figref}[1]{Fig. \ref{#1}}

\newcommand{\1}{_{(1)}}

\newcommand{\minus}{\scalebox{0.75}[1.0]{$-$}}
\newcommand{\inv}{^{\minus 1}}
\ncm{\vsir}{V_{SIR}}
\ncm{\Vsir}{V_{SIR}}
\ncm{\hsir}{H_{SIR}}

\ncm{\zit}[1]{\autocite{#1}}
\ncm{\GHS}{\textsc{HGS }}
\ncm{\Upot}{$U\!$-potential}
\ncm{\QUpot}{quasi-\Upot}
\ncm{\UE}{{V^E}}
\ncm{\VE}{{V^E}}
\ncm{\alpm}{\upsilon_\pm}
\ncm{\alp}{\upsilon_+}
\ncm{\alm}{\upsilon_-}
\ncm{\vpm}{\upsilon_\pm}
\ncm{\xpm}{x_\pm}
\ncm{\qpm}{q_\pm}

\ncm{\vp}{\upsilon_+}
\ncm{\vm}{\upsilon_-}
\ncm{\vO}{v_{\O}}
\ncm{\vN}{v_N}
\ncm{\vt}{v_\tau}
\ncm{\ut}{u_\tau}
\ncm{\apm}{a_\pm}
\ncm{\bpm}{b_\pm}
\ncm{\upm}{u_\pm}
\ncm{\qp}{q_+}
\ncm{\qc}{q_c}
\ncm{\up}{u_+}
\ncm{\xp}{x_+}
\ncm{\xc}{x_c}
\ncm{\epm}{\varepsilon_\pm}
\ncm{\fpm}{f_\pm}
\ncm{\Apm}{A_\pm}
\ncm{\Bpm}{B_\pm}
\ncm{\Dpm}{D_\pm}
\ncm{\Dp}{\vp}
\ncm{\Dc}{\Delta_c}
\ncm{\Spm}{S_\pm}
\ncm{\rSpm}{\rho S_\pm}
\ncm{\thpm}{\theta_\pm}

\ncm{\ntg}{\notag\\}

\ncm{\Ss}{S_{\textsl{sample}}}
\ncm{\Is}{I_{\textsl{sample}}}
\ncm{\Zs}{Z_{\textsl{sample}}}
\ncm{\Es}{E_{\textsl{sample}}}
\ncm{\Ns}{N_{\textsl{sample}}}
\ncm{\rhos}{\rho_{\textsl{sample}}}
\ncm{\gs}{\gamma_{\textsl{sample}}}
\ncm{\Zrge}{Z_{\rho,\gamma,E}}
\ncm{\Zmax}{Z_{\max}}
\ncm{\Qmax}{Q_{\max}}
\ncm{\el}{e^{\lambda}}
\ncm{\Ve}{V_{\epsilon}}
\ncm{\He}{H_{\epsilon}}

\ncm{\HPe}{H_{\PP,\epsilon}}
\ncm{\Emax}{E_{\max}}
\ncm{\rep}{\rho_{\epsilon}}
\ncm{\qe}{q_{\epsilon}}
\ncm{\expe}{\exp_{\epsilon}}
\ncm{\lne}{\ln_{\epsilon}}
\ncm{\Vpme}{V_{\pm,\ep}}
\ncm{\Vpe}{V_{+,\ep}}
\ncm{\Vme}{V_{-,\ep}}
\ncm{\qpme}{q_{\pm,\ep}}
\ncm{\qpe}{q_{+,\ep}}
\ncm{\qme}{q_{-,\ep}}
\ncm{\xpme}{x_{\pm,\ep}}
\ncm{\xpe}{x_{+,\ep}}
\ncm{\xme}{x_{-,\ep}}
\ncm{\qG}{q_\G}
\ncm{\pG}{p_\G}
\ncm{\ys}{y_2^*}
\ncm{\xs}{x_1^*}
\ncm{\vs}{v_2^*}
\ncm{\us}{u_1^*}
\ncm{\fl}{\varphi_\tau}
\ncm{\Gas}{\Gamma_\sigma}
\ncm{\tmp}{\age}
\ncm{\tImp}{\tau_{\I,\max}}

\begin{document}

%\begin{titlepage}
\title[Symmetries and normalization in 3-compartment epidemic models \RN{1}]
{Symmetries and normalization in 3-compartment epidemic models\\
\vspace{0.2 cm}
\RN{1}: The replacement number dynamics}
\author{Florian Nill}

%\address{Version, 22-Dec-2022}
\date{31-Dec-2022}

\thanks{The author is retired physicist, Dr.rer.nat.habil., formerly senior research fellow at Inst. theor. Physik, Freie Universität Berlin.
%Der Autor ist promovierter und habilitierter Physiker im Ruhestand.
}
\email{nill.florian@gmail.com}

%\dedication{Preliminary Version}

%\end{titlepage}

\begin{abstract}
%{\em Abstract still has to be revised}
% with linear vaccination and immunity waning rates are well known to show a transition from a  disease-free to an endemic equilibrium as the basic reproduction number is raised above threshold. 
As shown recently by the author, constant population SI(R)S models map to  Hethcote's classic endemic model originally proposed in 1973. This unifies a whole class of models with up to 10 parameters all being isomorphic to a simple 2-parameter master model for endemic bifurcation. In this work this procedure is extended to a 14-parameter {\em SSISS Model}, including social behavior parameters, a (diminished) susceptibility of the $R$-compartment and unbalanced constant per capita birth and death rates, thus covering many prominent models in the literature. Under mild conditions, in the dynamics for fractional variables in this model all vital parameters become redundant at the cost of possibly negative incidence rates. There is a symmetry group $G_S$ acting on parameter space $\A$, such that systems with $G_S$-equivalent parameters are isomorphic and map to the same normalized system. 
Using $(\Reff,I)$ as canonical coordinates, $\Reff$ the replacement number, normalization reduces to parameter space 
$\A/G_S$ with 5 parameters only.
This approach reveals unexpected relations between  various models in the literature. Part two of this work will analyze equilibria, stability and backward bifurcation  and part three will further reduce the number of essential parameters from 5 to 3.

% Members of such a compartment could be interpreted e.g. as ``recovered and partly immune'' or ``vaccinated with incomplete efficacy'' or  ``following lock-down measures'' or simply ``educated''. Switching on 
%$R$-susceptibility amounts to adding a third parameter to the normalized master model giving rise to a backward bifurcation as already observed by \autocite{Had_Cast} and later by many others.
\end{abstract}

\maketitle

\footnotesize
\tableofcontents
%	\newpage
\normalsize

%\begin{center}
%{\em All models are wrong, but some are useful}
%[George E.P.Box]
%\end{center}

\section{Introduction}

Building mathematical models to describe phenomena in natural sciences one typically encounters dynamical variables and external parameters. Within the model values for external parameters are considered to be given from outside, like fundamental natural constants (speed of light $c$, Planck's constant $\hbar$), parameters describing material or biological properties (spring constant $\kappa$, birth rate $\delta$, recovery rate $\ga$) or social behavior (contact rate $\be$). Naturally, reducing the number of essential parameters is always a goal to detect redundancies within parameter space and to simplify computations by unloading formulas. In the simplest case a pure dimensional scale parameter may without loss be put equal to one by choosing dimensional units appropriately. For example, putting $c=1$ amounts to measuring spatial distances by light running times and masses in units of energies, putting $\hbar=1$ amounts to measuring energies by angular frequencies and putting $\ga=1$ amounts to measuring time in units of the recovery time in an epidemic model. 

More generally a normalization program consists of finding appropriate coordinate transformations in variable+parameter space such that the transformed system only depends on a maximally reduced subset of transformed parameters. Examples are\footnote{The variables in these examples are:
\\
- Harmonic oscillator:  $u=q$, $v=p/\sqrt{mk}$, where $q,p,\kappa, m$ are coordinate, momentum, spring constant and particle mass and where the oscillation period is normalized to $T=2\pi$ by putting $m/k=1$.
\\
- Predator-prey model: $(u,v)$ denote appropriately rescaled prey and predator populations, respectively, and the predator mortality rate is normalized to one.
\\
- SIR model: $u=r_0S$, $v=r_0I$, where $r_0$ is the basic reproduction number, $(S,I)$ are susceptible and infectious fractions of the population and where the recovery rate is normalized to $\ga=1$.
\\
- Endemic model: $(u,v,r_0,\ga)$ as above, $c_1=\delta/(\ga+\delta)$ and $c_2=r_0c_1$, where $\delta$ is the balanced birth/mortality rate and where now time scale is normalized to $\ga+\delta=1$.}

\begin{equation}\label{normalization}
\begin{array}{rclrcl}
\multicolumn{3}{l}{\text{Harmonic oscillator}}\qquad\qquad &
\multicolumn{3}{l}{\text{Predator-prey model}}
\\
\dot{u} &=& v	&	\dot{u}&=&-uv+c_1u
\\
\dot{v} &=& -u	&	\dot{v}&=&uv -v
\\
\\
\multicolumn{3}{l}{\text{Classic SIR model}}\qquad\qquad &
\multicolumn{3}{l}{\text{Classic endemic model}}
\\
\dot{u} &=& -uv	&	\dot{u}&=&-uv-c_1u+c_2
\\
\dot{v} &=&uv-v	&	\dot{v}&=&uv -v
\end{array}
\end{equation}

Following this strategy the 6-parameter SI(R)S model ($\equiv$ combined SIRS/SIS model) with standard incidence, constant vaccination and immunity waning rates and a balanced birth and death rate has recently been shown by the author \autocite{Nill1} to admit a normalized version looking like the classic endemic model above\footnote{Aapart from allowing also values $u\in\RR$ and an enlarged parameter range
$(c_1,c_2)\in\RR_+\times\RR\,\cup\,\{0,0\}$.}.

\bsn
In this work (including two follow ups to be denoted as parts II and III \autocite{Nill3, Nill4}) this 
%By the methods of  this paper this result extends to a generalized 10-parameter SI(R)S model containing also an $I$-linear vaccination rate, a vaccination ratio for newborns and compartment dependent birth and mortality rates
%as long as the total population remains constant. The normalized model still looks the same as above, thus revealing a great redundancy in parameter space. In particular, this approach provides a unifying view on various results in the literature concerning equilibrium states, endemic bifurcation and stability properties for this class of models. Put differently, in the presence of a common normalized version  presenting basically repeated arguments for various subsets of non-vanishing parameters becomes obsolete. 
method is extended to the case where immunity after recovery (or vaccination) is incomplete right from the onset and where also compartment dependent constant per capita birth and death rates lead to a time varying population size $N$. In this way one is naturally lead to replacing the SI(R)S model by a 
{\em SSISS model}, where in place of the usual $S$, $I$ and $R$ compartments we have two susceptible compartments $\SS_1$ and $\SS_2$ and one infectious compartment $\II$. Infection transmission from $\II$ to $\SS_2$ is diminished as compared to transmission to $\SS_1$. There is a vaccination flow from $\SS_1$ to $\SS_2$ and an immunity waning flow from $\SS_2$ to $\SS_1$. 
The model could also be interpreted by considering $\SS_2$ as the  ``lock-down'' fraction and $\SS_1$ as the ``freedom fraction''. In this picture flows from $\SS_1$ to $\SS_2$ and vice-versa are described by an $\II$-linear (respectively $(N-\II)$-linear) flow with rate parameters $\theta_i,\ i=1,2$, modeling social behavior in reaction to published prevalence data. Combining both interpretations it turns out to be convenient to start with an abstract version of a SSISS model staying completely symmetric under interchanging $\SS_1$ and $\SS_2$, see Fig. \ref{Fig_SSI-Flow}.

\bsn
The present part I provides a normalization prescription reducing the number of independent parameters in this model from initially fourteen to essentially five (four in the SI(R)S model sub-case). 
Based on this approach, part II will give  a complete review on equilibria and stability in the master SSISS model, thereby also recovering an exceptional scenario which had been overlooked  in the literature so far.
In part III the scaling symmetry for  SI(R)S models mentioned above will be generalized to the full SSISS model, thereby reducing the number of parameters again by two. 
So, the total reduction from fourteen to three reveals a great hidden redundancy in parameter space. It also provides a unifying view on results in the literature concerning equilibrium states, endemic bifurcation and stability properties for all kinds of sub-classes of this model. Put differently, in the presence of a common normalized version  presenting basically repeated arguments for various subsets of non-vanishing parameters becomes obsolete. 

\bsn
Relating this work to the literature, let me focus on deterministic SIR-type  3-compartment dynamical systems, which conveniently may be classified  according to

\begin{itemize}
\item[A)]
constant vs. time-varying total population size $N$,
\item[B)]
infection transmission only from $I$ to $S$ vs. also from $I$ to $R$ (in which case it makes sense to rename 
$S\equiv S_1$ and $R\equiv S_2$). 
\end{itemize}
Also, I will restrict this survey to models with standard bi-linear incidence flows $\be_i\SS_i\II/N$, such that the vector field 
$\dot{\bY}=\bV(\bY)$, $\bY=(\SS_1, \SS_2,\II)$, is homogeneous of first order. This applies to diseases where the number of effective contacts per capita is independent of $N$.

\ssn
{\bf ad A)} Endemic models with constant population have first been constructed by adding a non-zero balanced birth and death rate to the classic SIR model of \autocite{KerMcKen}. As shown 
by \autocite{Hethcote1974} (see also
\autocite{Hethcote1976, Hethcote1989}), in this way already the simplest model without vaccination and loss of immunity shows a bifurcation from a stable disease-free equilibrium point (DFE) to a stable endemic scenario when raising the basic reproduction  number $R_0$ above one. Nowadays this is considered as Hethcote's {\em classic endemic model}.
Including linear vaccination and/or loss of immunity terms and optionally also considering recovery without immunity  one ends up with various types of constant population SI(R)S models without changing this picture, see for example \autocite{KorobWake, ORegan_et_al, Chauhan_et_al, Batistela_et_al}. As remarked above (and reviewed in more detail in Appendix \ref{Sec_SIRS}), the true reason lies in the fact that constant population SI(R)S models with up to 10 parameters all map to the same normalized 2-parameter version of the classic endemic model as given in \Eqref{normalization}.

\ssn
Models with variable population are mostly studied under the assumption of a constant (i.e. $N$-independent) birth flow. Heuristically this may be justified by assuming that $N$ varies slowly on  characteristic epidemic time scales. But truly speaking, as already pointed out by \autocite{Mena-LorcaHeth}, this Ansatz rather models a constant immigration scenario.
So in this work I will follow the  more natural proposal of modeling vital dynamics by possibly department dependent constant per capita birth and death rates. 
Note that, unless fine tuning parameters, this implies that either $N(t)\rto\infty$ or $N(t)\rto 0$ as $t\rto\infty$. So in this type of models one always analyzes the dynamics of fractional variables 
$S_i:=\mathbb{S}_i/N,\,I:=\II/N$, which  is well known to be independent of $N(t)$.
Apparently, this stream of models has been initiated by \autocite{BusDries90,
BusDries91, DerrickDriessche}. 
\autocite{Razvan} has studied a SIRS model in this sense with infection transmission also from outside and a SIS-version  with  varying population size has been analyzed by 
\autocite{LiMa2002}. For generalizations to SEIR models see e.g.  \autocite{Greenhalgh,LiGraLiKa,SunHsieh,LuLu}.

\ssn
{\bf ad B)} 
A different approach to modeling partial and/or waning immunity consists of introducing a diminished incidence flow with rate 
$\be_R\equiv\be_2>0$ directly from $R\equiv S_2$ to $I$. This has presumably first been proposed in the so-called SIRI model of \autocite{DerrickDriessche}, see above. In addition, the authors also introduced a time varying population size $N(t)$ and an excess mortality $\Del\mu_I$ in compartment $I$ to this model. In turn, they didn't use linear vaccination nor immunity waning terms. In this way they identified a range of parameters in the domain $R_0<1$, for which besides the locally asymptotically stable disease free equilibrium there also coexist two endemic equilibria, one being a saddle and the other one also being  locally asymptotically stable.
Later \autocite{Had_Cast} found the same phenomenon in their combined SIS/SIRS core group model with linear vaccination, constant population and also two incidence rates $\be_i$ for $S\rto I$ and $R\rto I$. Meanwhile it is well known that models with infection incidents from several compartments may show a so-called {\em backward bifurcation} from the disease-free to an endemic scenario \autocite{Had_Dries}. This means that two locally asymptotically stable  equilibrium states may coexist for some range below threshold, causing also hysteresis effects upon varying parameters. Apparently, a varying population size is not needed for this. In 
\autocite{KribsVel} the authors have improved and extended these results by adding also a linear immunity waning rate to  the model of \autocite{Had_Dries}. 

One may also distinguish vaccinated and recovered people into separate compartments. This leads to 4-compartment models, where similar results have been obtained by, e.g.  \autocite{Arino_et_al, Yang_et_al}. 

Backward bifurcation has lately also been observed in SEIRS-type models for Covid-19 by considering two distinguished susceptible compartments. In \autocite{NadimChatto} the less susceptible compartment had been interpreted as an incomplete lockdown and in \autocite{Diagne_et_al} as an incomplete vaccination efficacy. 

More recently, in \autocite{AvramAdenane2022, AvramAdenane_et_al} the authors have given a thorough stability analysis of an eight parameter SIRS-type model by adding a varying population size to the model of \autocite{KribsVel} (apparently without being aware of that paper).

Closing this overview I should also remark that backward bifurcation is also observed when considering $I$-dependent contact or recovery rates to model reactive behavior or infection treatment. However the list of papers on this topic over the last 20 years becomes too huge to be quoted at this place. 

\bsn
This paper extends the normalization algorithm for constant population SI(R)S models to models as above, i.e. with time varying population size and/or a non-zero incidence rate 
$\be_R\equiv\be_2$ from $R\equiv S_2$ to $I$.
As a starting observation, there is an ambiguity in deriving the dynamics $\dot{\by}=\bF(\by)$ for fractional variables 
$\by=(S_1,S_2,I)$, see Appendix \ref{Sec_vital-dynamics}. This allows choosing the vector field $\bF$ such that all vital dynamics parameters become redundant, provided the birth-minus-death rates $\nu_i=\delta_i-\mu_i$ in $S_1$ and $S_2$ coincide, $\nu_1=\nu_2=\nu$.
This redundancy already reduces the number of parameters in the master SSISS model from fourteen to eight. More than that, $\bF$ depends on the incidence rates $\be_i$ only as a function of 
$\tilde{\be}_i=\be_i+\nu_I-\nu$, where $\nu_I=\delta_I-\mu_I$ is the birth-minus-death rate in $I$. Assuming for simplicity compartment independent birth rates gives
$\tilde{\be}_i=\be_i-\Del\mu_I$, where $\Del\mu_I$ denotes the excess mortality in $I$. In this way models with variable population, 
$\Del\mu_I>0$, and absence of a incidence rate from $R$, 
$\be_2=0$, look like models with constant population, $\Del\mu_I=0$, and a {\em negative incidence rate $\be_2=\tilde{\be}_2<0$}. Conversely, models with positive incidence rates $\be_i>0$ and excess mortality $\Del\mu_I<\min\{\be_1,\be_2\}$ behave like models with constant population size and incidence rates $\be_i=\tilde{\be_i}>0$. So, the above classification schemes A) and B) become blurred and, instead,  it is more expedient to view all models as if they had constant population size and two distinguished and possibly also negative incidence rates 
$\tilde{\be}_i\in\RR$.

\bsn
In this way most of the above bench marking 3-compartment models (if necessary after imposing the constraint $\nu_1=\nu_2$) become comparable as sub-cases of the master SISS model, with tilde parameters swallowing all birth and death rates and possibly with negative incidence rates 
$\tilde{\be_i}\in\RR$. 
As an example, the models of \autocite{Had_Cast} and \autocite{KribsVel} become isomorphic and they completely cover the sub-case 
$\mu_1=\mu_2$ and $0<\min\{\tilde{\be}_1,\tilde{\be}_2\}$
in \autocite{AvramAdenane2022}. Also, apart from an irrelevant boundary case, the complementary  sub-case  
$\mu_1=\mu_2$ and $0>\min\{\tilde{\be}_1,\tilde{\be}_2\}$ 
in \autocite{AvramAdenane2022}
is covered by the model of \autocite{LiMa2002}. So, applying the normalization procedure of this paper, all results in 
Section 5 and 6 of \autocite{AvramAdenane2022} already follow from  the previous literature.
A more detailed list of unexpected relations between the above models is given in Section \ref{Sec_examples}.

\bsn
The plan of this paper is as follows. In Sections \ref{Sec_const-pop} and \ref{Sec_time-dependent} we pass to fractional compartment variables, $S_i=\SS_i/N$ and $I=\II/N$, and prove redundancy of all vital dynamics parameters at the cost of possibly negative 
%become redundant. Due to an ambiguity in the definition of the vector field 
%$\dot{\by}=\bF(\by)$, $\by=(S_1,S_2,I)$, this fact is mostly overlooked in the literature\footnote{Appendix \ref{Sec_vital-dynamics} provides a general algorithm showing how to get this result in any $n$-compartment model where the population size $N$ only changes via constant per capita birth and death rates, 
%$\dot{N}/N=\sum_i\nu_iy_i$.
%}.
%The price to pay is that one possibly has to allow negative effective 
incidence rates $\tilde{\be}_i$. For convenience, time scale is also normalized by putting the total expected waiting time in compartment $I$ equal to one. In this way the number of essential parameters is already reduced from fourteen to seven. Thus, denoting $\A$ the space of essential parameters, we have 
$\dim\A=7$.

Section \ref{Sec_par_space} classifies various useful subsets in parameter space like 
$\Aph\subset\A$, guaranteeing forward invariance of the 
{\em physical triangle} 
$$
\Tph:=\{(S_1,S_2,I)\in\RRN^3\mid S_1+S_2+I=1\},
$$
and $\Abio\subset\Aph$, guaranteeing an epidemiological interpretation of parameters by requiring in particular 
$\theta_1\geq 0\geq \theta_2$.

Section \ref{Sec_examples} identifies eight examples from the above list of models as sub-cases of the master SSISS model. In this way we obtain various relations between these models as indicated above, which apparently have not been recognized before.

In Section \ref{Sec_periodic-sol} we adapt methods from 
\autocite{BusDries90} to prove absence of periodic solutions for all parameters non-negative, except $\be_i$. The extension to parameters $\bfa\in\Abio$ (requiring $\theta_2\leq 0$) heavily relies on the symmetry results in Section \ref{Sec_Repl.no} and will be proven in Section \ref{Sec_results}. 

\bsn
Section \ref{Sec_Repl.no} starts from the observation, that the time-normalized equation of motion for $I$ takes the generic form $\dot{I}=(\Reff-1)I$, where  
$\Reff=\be_1S_1+\be_2S_2$
is  the {\em replacement number} \autocite{Hethcote2000},  i.e. the expected number of secondary cases produced by a typical infectious individual during its time of infectiousness (nowadays mostly called {\em effective reproduction number}). 
A coordinate free formulation of the model naturally leads to taking 
$(\Reff,I)$  as independent {\em canonical coordinates}\footnote{Here ``canonical'' is not meant in the sense of Hamiltonian systems.}
in the physical triangle $\Tph$.
In this way, we arrive at formulating the SSISS model as a dynamical system in $(\Reff,I)$-space, called the {\em replacement number (RN) dynamics} (Section \ref{Sec_Can-Coord}).
\begin{equation}
\dot{X}_{\mathrm{rep}}=f(\Reff,I),\qquad \dot{I}=(\Reff-1)I.
\label{RN-system}
\end{equation}
Since $f(\Reff,I)$ turns out to be a 5-parameter quadratic polynomial with no term $\sim\Reff^2$, the number of free parameters is now reduced from seven to five.

%More precisely, denoting $\D$ the space of these new parameters, 
%$\dim\D=5$, mapping the SSISS model to the RN-dynamical system \eqref{RN-system} above will induce a diffeomorphism between parameter spaces $\phi:\A\ni\bfa\mapsto(\bx(\bfa),\be_1,\be_2)\in\D\times\B$, where 
%$\B:=\{(\be_1,\be_2)\in\RR^2\mid\be_1>\be_2\}$ and where the polynomial $f(\Reff,I)$ only depends on 
%$\bx(\bfa)\in\D$. Hence, the system \eqref{RN-system} becomes independent of the incidence parameters $\be_i$ and their role is reduced to fixing the image of the physical triangle in coordinates  $(\Reff,I)$.

\bsn
The main results of this paper are derived in Section \ref{Sec_results}.
Denoting $\D$ the new parameter set, $\dim\D=5$, the above approach yields a surjective submersion $\A\ni\bfa\mapsto\bx(\bfa)\in\D$. Moreover, $\A$ becomes a principal fibre bundle with respect to a group right action 
$\li:\A\times G_S\rto\A$ such that $\bx(\bfa\li\bg)=\bx(\bfa)$ and $\D\cong\A/G_S$. Here $G_S\subset GL_+(\RR^2)$ is the group acting on $(S_1,S_2)\in\RR^2$ and leaving $S_1+S_2$ invariant.
\Eqref{RN-system} implies that SSISS dynamical systems at parameter values $\bfa,\bfa'\in\A$ are isomorphic whenever $\bfa$ and $\bfa'$ are $G_S$-equivalent, i.e.
$\bx(\bfa)=\bx(\bfa')$ or equivalently $\bfa'=\bfa\li\bg$ for some $\bg\in G_S$. In this way we also get 
\begin{itemize}
\item[-]
Absence of periodic solutions also for parameters $\bfa\in\Abio$,
\item[-]
Conditions under which the social behavior parameters $\theta_i$ can be ``gauged to zero'', i.e. there exists $\bg\in G_S$ such that $\bfa\li\bg\in\A_{\bth=\bzero}$.
\end{itemize}  
Section \ref{Sec_examples2} revisits the examples from the literature within the new formalism and Section \ref{Sec_Summary} gives a summary and outlook to parts II and III of this work.
Finally, Appendix \ref{Sec_vital-dynamics} provides a normalization prescription for the dynamics of fractional variables in $n$-compartment models with linear (i.e. constant per capita) birth and death rates, Appendix \ref{Sec_SIRS} reviews the scaling symmetry in SI(R)S models introduced in \autocite{Nill1} and Appendix \ref{Sec_a=0} discusses a boundary case in parameter space.

\section{The SSISS model\label{Sec_SSISS-model} }
\begin{figure}[ht!]
\centering
\includegraphics[width=0.8\textwidth]
	{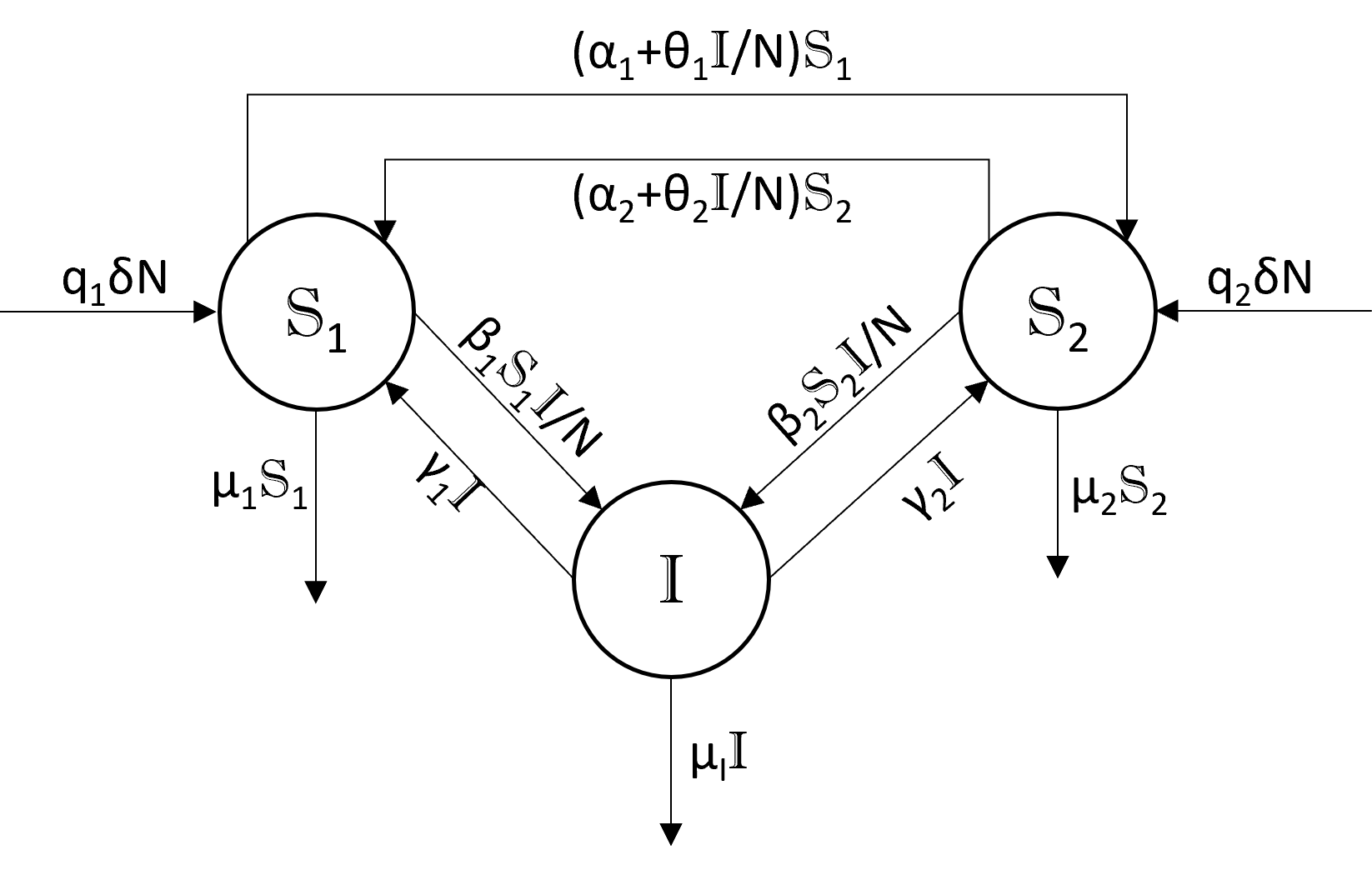}
	\caption{Completely symmetric flow diagram of the SSISS model. 
	All parameters are nonnegative except 
	$\theta_2\in[-\al_2,0]$. Also $q_1+q_2=1$, $\ga_1+\ga_2>0$ and $\beta_1>\beta_2$. Generalizing to compartment dependent birth rates amounts to replacing $\delta N$ by 
	$\delta_1\SS_1+\delta_2\SS_2+\delta_I\II$.}
	\label{Fig_SSI-Flow}
\end{figure}

This Section starts with proposing an abstract completely symmetrized SSISS model consisting of three compartments, $\mathbb{S}_1$, $\mathbb{S}_2$ and $\mathbb{I}$, with total population $N=\mathbb{S}_1+\mathbb{S}_2+\mathbb{I}$. Members of $\mathbb{I}$ are infectious, members of $\mathbb{S}_1$ are highly susceptible (socially active or not immune) and members of $\mathbb{S}_2$ are less susceptible (partly immune or reducing contacts). The flow diagram between compartments is depicted in Fig. \ref{Fig_SSI-Flow}.

The parameters in this model may be given the following interpretations

\begin{tabular}{lcp{0.75\textwidth}}
$\al_1$ &:& 
		Vaccination rate of susceptibles moving from 
		$\mathbb{S}_1\rto \mathbb{S}_2$
		(assuming $\theta_1=\theta_2=0$, see below).
\\
$\al_2$ &:& 
		Immunity waning rate inducing a flow from 
		$\mathbb{S}_2\rto \mathbb{S}_1$ (assuming $\theta_2=0$, see below).
\\
$\be_i$ &:& 
		Number of effective contacts per unit time of a 
		susceptible from $\mathbb{S}_i$.
\\
$\ga_i$ &:& 
		Recovery rate from $\mathbb{I}\rto \mathbb{S}_i$.
\\
$\theta_1$ &:& 
		Willingness to get vaccinated (alternatively 
		to reduce contacts) given the actual prevalence 
		$\mathbb{I}/N$. In reality only one of the two parameters 
		$\al_1$ and $\theta_1$ should be chosen non-zero.
\\
$\theta_2$ &:& 
		Epidemiologically one should restrict to $\theta_2=0$ or
		($\theta_2=-\al_2<0$ and $\al_1=0$). In this latter
		case the meaning of the $\mathbb{S}_2$-compartment is 
		``contact reducing'' and 
		$\al_2=-\theta_2$ parametrizes the readiness to 
		increase contacts proportional to $1-\mathbb{I}/N$.
\\
$\mu_i$ &:& 
		Mortality rate in $\mathbb{S}_i$.
\\
$\mu_I$ &:& 
		Mortality rate in $\mathbb{I}$. One could also consider 
		vertical transmission, in which case $\mu_I$ would be 
		the mortality rate diminished by the rate of infected 
		newborns.
\\
$\Del\mu_I$	&:&
		Mortality excess $\Del\mu_I=\mu_I-\mu$ in case 
		$\mu_1=\mu_2=\mu$, which will be assumed most of the time.
\\
$\delta$ &:& 
		Rate of not infected newborns. Generalizing to compartment dependent birth rates amounts to replacing 
		$\delta N=\delta_1\SS_1+\delta_2\SS_2+\delta_I\II$. 
\\
$q_i$  &:& 
		Split ratio of newborns between $\mathbb{S}_1$ and $\mathbb{S}_2$, 
		$q_1+q_2=1$. In the reduced-immunity interpretation 
		$q_2$ would be the portion of vaccinated newborns.
\end{tabular}

\bsn
So in total this model counts 15 independent parameters (12 if we require constant total population, $\delta_i=\mu_i,\ \delta_I=\mu_I$). Epidemiologically all parameters except 
$0\geq\theta_2\geq-\al_2$ are assumed non-negative and also 
$\be_2<\be_1$. A more technical classification of admissible parameter ranges will be given below. 
Here is a list of prominent examples in the literature 

\begin{itemize}
\item[-]
Hethcotes classic 3-parameter endemic model 
\autocite{Hethcote1974, Hethcote1976, Hethcote1989} by putting $\delta=\mu_i=\mu_I>0$, $q_1=1$, $\beta_1>0$, 
$\ga_2>0$ and all other parameters vanishing.
\item[-]
The 7-parameter SIRS model with time varying population size in \autocite{BusDries90}, adding to Hethcote's model  an immunity waning rate $\al_2$ and allowing different (constant per capita) mortality and birth rates.
\item[-]
The 6-parameter SIRI model of \autocite{DerrickDriessche}, replacing the immunity waning rate $\al_2$ in \autocite{BusDries90} by the incidence rate $\be_2>0$ and also requiring $\mu_1=\mu_2$.
\item[-]
An extended 10-parameter constant population SI(R)S (i.e. mixed SIRS/SIS) model with  constant and $I$-linear vaccination rates 
$\al_1, \theta_1$, an immunity waning rate $\al_2$ and two recovery flows $I\leftarrow S_i$. Hence $\delta_i=\mu_i,\ \delta_I=\mu_I$ and $\theta_2=\beta_2=0$\footnote{Here I have chosen enlarge the conventional setting for SI(R)S models by also allowing $\theta_1>0$.}.
\item[-]
The 6-parameter isolated core system in \autocite{Had_Cast}, with  two incidence and recovery rates, $\be_i, \ga_i>0$, a vaccination term $\al_1>0$ and a constant population with balanced birth and death rates, $\delta=\mu_i=\mu_I>0$ and $q_1=1$. 
\item[-]
The 7-parameter vaccination models of  \autocite{KribsVel} adding an immunity waning rate $\al_2>0$ to the model of \autocite{Had_Cast}. As we will see in \Eqref{KZVH} below, due to a redundancy of parameters the two models actually stay isomorphic.
\item[-]
The 8-parameter SIS-model with vaccination and varying population size of \autocite{LiMa2002} keeping only 
$\theta_i=\ga_2=\be_2=0$ and assuming $\mu_1=\mu_2=\mu$.\footnote{Actually the authors let $\mu$ be a function of $N$, which however disappears when passing to fractional variables.}
As we will see in \eqref{LM}, after a parameter transformation this model becomes isomorphic to the case where only $\theta_i=0$ and $\be_2\leq 0$.
\item[-]
The 8-parameter SIRS-type model analyzed recently by \autocite{AvramAdenane2022}, keeping only 
$\ga_1=\theta_1=\theta_2=q_2=0$ 
and  all other parameters positive.
The authors allow a varying population size by first discussing  the general case of all mortality rates being different and then concentrate on $\mu_1=\mu_2\neq \delta$ and $\Del\mu_I>0$. Their paper is closest to the present work and in fact initiated it.
%\footnote
%{For convenience of the reader Appendix \ref{Avram_parameters} provides a table of parameter symbols used in the present paper, mapping them to the ones chosen in \autocite{AvramAdenane2022} and \autocite{KribsVel}, respectively.}.
%\item[-]
%{\em Model-1} of this paper generalizing the cases of
%\autocite{Had_Cast, KribsVel} by keeping $N=\const$ and 
%$\theta_2=0$ as the only restriction. 
%Actually, when modeling inefficient immunity as a reduced susceptibility by introducing $\be_2>0$, the additional  parameter $\al_2$  governing immunity waning may also safely be dropped without changing qualitative results.\footnote
%{In fact, as we will see in Section 
%\ref{Sec_Can-Coord}, the dynamics of the model  depends on $\al_i$
%only in the combination 
%$a:=\al_1+\al_2$ and $R_0:=(\al_1\be_2+\al_2\be_1)/a$. So dropping $\al_2$ lowers the threshold  $R_0$ but doesn't change the overall scenario.}
%As we will see in Theorem
%\ref{Thm_N_varying}, for $\mu_1=\mu_2$,
%$\Del\mu_I<\be_2$ and $\theta_i=0$ Model-1 is also isomorphic to the case of time dependent $N(t)$ studied by \autocite{AvramAdenane2022}.
%\item[-]
%{\em Model-2} of this paper by putting $N=\const$, 
%$\al_1=0$ and $\al_2=-\theta_2>0$. So in place of vaccination vs. loss of immunity this model describes social contact behavior.
\end{itemize} 

\bsn
In a ``zeroth normalization'' step I will now show that passing to fractional variables and requiring  
$\delta_1-\mu_1=\delta_2-\mu_2$ all vital dynamic parameters in the SSISS model become redundant\footnote{
Redundancy of constant per capita birth and death rates may in fact be shown under quite general assumptions in $n$-compartment models, see Appendix \ref{Sec_vital-dynamics}.}. 
In this way the number of essential parameters reduces from 14 to 8. The price to pay in the non-constant population case is possibly getting negative incidence rates 
$\be_i$.

\subsection{Constant population\label{Sec_const-pop}}

To get a constant population $N$ the birth rates have to obey 
$\delta_i=\mu_i$ and $\delta_I=\mu_I$, or more generally
\begin{equation}
\delta=(\mu_1 \mathbb{S}_1 +\mu_2 \mathbb{S}_2 + \mu_I \mathbb{I})/N\,.
\label{delta} 
\end{equation}
In case $\mu_1=\mu_2=\mu$ this would read
$\delta=\mu +I\Del\mu_I $. Heuristically this should be understood as an approximation for $\Del\mu_I/\mu\ll 1$. Under this assumption, denoting fractions of the total population by $S_i=\mathbb{S}_i/N$ and $I=\mathbb{I}/N$ and introducing the notations 
\begin{equation}
\begin{array}{rclrcl}
\tal_1 &:=&\al_1+q_2\mu_1\,,\qquad\qquad
& \tga_1 &:=&\ga_1+q_1\mu_I\,,
\\
\tal_2 &:=&\al_2+q_1\mu_2\,,& \tga_2 &:=&\ga_2+q_2\mu_I\,,
\end{array}
\label{notation} 
\end{equation}

\begin{equation}
\bS=\begin{pmatrix}
S_1 \\
S_2 
\end{pmatrix},\qquad
\bD(\bbe)=\begin{pmatrix}
\beta_1 & 0\\
0&\beta_2
\end{pmatrix},
\quad
\bE(\bal)=\begin{pmatrix}
\al_1 & -\al_2\\
-\al_1 &\al_2
\end{pmatrix},
\quad
\tilde{\bga}=\begin{pmatrix}
\tga_1 \\
\tga_2 
\end{pmatrix}
\label{matrix} 
\end{equation}
the dynamical system described by the flow diagram Fig. \ref{Fig_SSI-Flow} becomes
\begin{eqnarray}
\dot{\bS} &=& 
	-\left[
	\bE(\tilde{\bal})+I\bE(\bth)+I\bD(\bbe)
	\right]	
	\bS+I\tilde{\bga}\,,
\label{dot_S}\\
\dot{I} &=&\tilde{\ga}(\Reff-1)I\,,
\qquad\tilde{\ga}=\tilde{\ga}_1+\tilde{\ga}_2
\label{dot_I} \\
\Reff &:=& (\beta_1S_1+\beta_2S_2)/\tilde{\ga}\,.
\label{Replacement-Nr} 
\end{eqnarray}
Note that $\tilde{\ga}\inv\equiv(\ga_1+\ga_2+\mu_I)\inv$ is the expected waiting time in $I$ and hence $\Reff$ is the
{\em replacement number} \autocite{Hethcote2000}, i.e. the expected number of secondary cases produced by a typical infectious individual during its time of infectiousness.  In conventional SI(R)S models, i.e. for 
$\beta_2=\theta_2=0$,
the replacement number in the limit $S_1=1$ would become the {\em basic reproduction number} $r_0=\beta_1/\ga$. This is why nowadays the replacement number is mostly called {\em effective reproduction number}. Later we will also have the notion of a {\em reduced reproduction number $R_0$} as the value of 
$\Reff$ at the disease-free equilibrium. To avoid misunderstandings, I prefer to keep the various notions of ``reproduction numbers'' for parameters, whereas the replacement number $\Reff$ is considered as a dynamical variable.

\bsn
Now obviously, by \eqref{notation}, all vital dynamics parameters become redundant and may be absorbed by redefining $\al_i$ and $\ga_i$. Note that this observation is independent of the choice of $\beta_i$ and $\theta_i$, i.e. it already holds in a combined SI(R)S model.

\subsection{Time varying population\label{Sec_time-dependent}}
To derive the equations of motion in case of a time varying population keep compartment dependent per capita birth and death rates 
$\delta_i,\delta_I, \mu_i, \mu_I$ constant and put 
$\bY=(\SS_1,\SS_2,\II)$, $\by=N\inv\bY$ and 
$$
\bnu\equiv(\nu_1,\nu_2,\nu_I)
:=(\delta_1-\mu_1,\delta_2-\mu_2,\delta_I-\mu_I).
$$ 
Then $\dot{\by}=\dot{\bY}/N-\by\dot{N}/N$ and 
$\dot{N}/N=\bra\bnu\mid\by\ket$. Using $S_1+S_2+I=1$ we may rewrite
\begin{align*}
%\dot{N}/N &=
%(\delta_1-\mu_1)S_1+(\delta_2-\mu_2)S_2+(\delta_I-\mu_I )I
%\label{dot_N}\\
S_1\dot{N}/N	&=S_1[\nu_1+(\nu_2-\nu_1)S_2+(\nu_I-\nu_1)I]
	\\
S_2\dot{N}/N	&=S_2[\nu_2+(\nu_1-\nu_2)S_1+(\nu_I-\nu_2)I]
	\\
I\dot{N}/N	&=I[\nu_I+(\nu_1-\nu_I)S_1+(\nu_2-\nu_I)S_2].
\end{align*}
So now introduce 
\begin{equation}
\begin{array}{rclrcl}
\tal_1 &:=&\al_1+q_2\delta_1\,,\qquad\qquad &
\tal_2 &:=&\al_2+q_1\delta_2\,, 
\\
\tga_1 &:=&\ga_1+q_1\delta_I\,,& \tga_2 &:=&\ga_2+q_2\delta_I\,,
\\
\tilde{\be}_1 &:=&\be_1+\nu_I-\nu_1\,,& 
\tilde{\be}_2 &:=&\be_2+\nu_I-\nu_2\,.
\end{array}
\label{notation2} 
\end{equation}
With the same notation as in \Eqref{matrix} and 
$\bfe(\bm{\nu}):=
\begin{pmatrix}
\nu_1-\nu_2 \\
\nu_2-\nu_1
\end{pmatrix}
$ 
we then get
\begin{eqnarray}
\dot{\bS} &=& 
	-\left[
	\bE(\tilde{\bal})+I\bE(\bth)+I\bD(\tilde{\bbe})
	\right]	
	\bS+I\tilde{\bga}+S_1S_2\bfe(\bm{\nu})\,,
\label{dot_S_N(t)}\\
\dot{I} &=&\tilde{\ga}(\Reff-1)I\,,
\label{dot_I_N(t)} \\
\Reff &:=& (\tilde{\beta}_1S_1+\tilde{\beta}_2S_2)/\tilde{\ga}\,,
\qquad\tilde{\ga}:=\tilde{\ga}_1+\tilde{\ga}_2.
\label{Replacement-Nr_N(t)} 
\end{eqnarray}
So, imposing the condition $\nu_1=\nu_2=:\nu$ and putting 
$\Del\nu_I:=\nu-\nu_I$ we get $\bfe(\bm{\nu})=0$ and the equations of motion look exactly as in the case of constant population \eqref{dot_S}-\eqref{Replacement-Nr}. Again all vital dynamics parameters become redundant and may be absorbed by redefining 
$\be_i$, $\al_i$ and $\ga_i$. The difference this time is that $\tilde{\be}_i=\be_i-\Del\nu_I$ may become negative! Thus we arrive at 

{\proposition\label{Prop_N_varying} 
Assume $\nu_1=\nu_2$. 
\begin{itemize}
\item[i)]
If $\Del\nu_I\leq\min\{\be_1,\be_2\}$ the SSISS model with variable population maps to the model with constant population. 
\item[ii)]
If $\Del\nu_I>\min\{\be_1,\be_2\}$ it maps to the model with 
$\min\{\be_1,\be_2\}=0$ and variable population with 
$\widetilde{\Del\nu_I}=\Del\nu_I-\min\{\be_1,\be_2\}$.
\item[iii)]
If $\Del\nu_I=\be_2<\be_1$ and $\theta_2=0$ it becomes the extended SI(R)S model with $\theta_1\geq 0$ and two recovery flows $I\rto S_1$ and $I\rto S_2$.
\end{itemize} 
}

{\remark
Note that under the usual assumptions $\delta_i=\delta_I=\delta$ and $\mu_1=\mu_2=\mu$, $\Del\nu_I$ coincides with the excess mortality in the infectious compartment, 
$\Del\nu_I=\mu_I-\mu=\Del\mu_I$. 
}

{\remark\label{Rem_neg-beta}
The observation that on the level of fractional variables in both scenarios (constant vs. variable population, the latter provided $\nu_1=\nu_2$) all vital dynamics parameters are redundant seems to be new\footnote
{As communicated privately this had also been realized recently in a talk by Florin Avram.}.
Essential for this is allowing all four parameters 
$(\al_i,\ga_i)$ being positive and $\be_i$ possibly being negative. The introduction 
of parameters $\theta_i$ is not needed to assure this. Redundancy of constant per capita birth and death rates may in fact be shown under quite general assumptions in $n$-compartment models, see Appendix \ref{Sec_vital-dynamics}.}

\subsection{Classifying parameter space\label{Sec_par_space}}
In this subsection assume $\nu_1=\nu_2$. Then the reformulation  in terms of possibly negative incidence rates 
$\tbe_i$ leads to a new classification scheme  identifying {\em seven sectors} in this model. For $\theta_i= 0$ these are labeled 
by the signatures of $\tilde{\beta}_1+\tilde{\beta}_2$ and 
$\tilde{\beta}_1\tilde{\beta}_2$ (in case of a compartment independent birth rate $\delta$ equivalently by the size of the excess mortality $\Del\mu_I$), see Table \ref{Tab_1}. For 
$\theta_i\neq 0$ this classification will be refined  in Section \ref{Sec_Repl.no}, Table \ref{Tab_1'}. 

\begin{table}[htpb!]
\caption{Seven sectors in the SSISS-model at $\theta_i=0$ and for compartment independent birth rate $\delta$. By Corollary \ref{Cor_Avr=JiLi+KrZa} Sector \RN{1} is isomorphic to the models of \autocite{KribsVel, Had_Cast} and Sectors \RN{3}-\RN{7} are largely covered by \autocite{LiMa2002}. Sector \RN{2} is a mixed SI(R)S model with two recovery flows $I\rto R$ and $I\rto S$.}
\label{Tab_1} 
\renewcommand{\arraystretch}{1.4}
\begin{tabular}{|c|c|c|l|l|}
\hline
	Sector & $\sign(\tilde{\be}_1+\tilde{\be}_2)$ & 
	$\sign(\tilde{\be}_1\tilde{\be}_2)$ & 
	Interval $[\tilde{\be}_1,\tilde{\be}_2]$ & 
	Excess mortality $\Del\mu_I$
\\
\hline
\RN{1}&$+$& $+$& $0<\tbe_2<\tbe_1$ & $\Del\mu_I<\be_2$ 
\\\hline
\RN{2} (SIRS)&$+$& $0$& $0=\tbe_2<\tbe_1$ & $\Del\mu_I=\be_2$ 
\\\hline
\RN{3}&$+$& $-$& $0<-\tbe_2<\tbe_1$ & $\be_2<\Del\mu_I<(\be_1+\be_2)/2$ 
\\\hline
\RN{4}&$0$& $-$& $0<-\tbe_2=\tbe_1$ & $\Del\mu_I=(\be_1+\be_2)/2$ 
\\\hline
\RN{5}&$-$& $-$& $\tbe_2<-\tbe_1<0$ & $(\be_1+\be_2)/2<\Del\mu_I<\be_1$ 
\\\hline
\RN{6}&$-$& $0$& $\tbe_2<\tbe_1=0$ & $\be_1=\Del\mu_I$ 
\\\hline
\RN{7}&$-$& $+$& $\tbe_2<\tbe_1<0$ & $\be_1<\Del\mu_I$ 
\\\hline
\end{tabular}
\renewcommand{\arraystretch}{1}
\end{table}

\bsn
To simplify notation, in what follows let me drop the tilde above parameters. The case 
$\be_1=\be_2$ will be ignored, since in this case putting $S=S_1+S_2$ one easily checks that 
$(S,I)$ obeys the dynamics of a SIS model, which can immediately be solved by separation of variables.  Also, due to the permutation symmetry $1\leftrightarrow 2$, there is no loss assuming $\beta_1>\beta_2$. 
Next, choosing time scale to be measured in units of 
$\ga\inv$, we may without loss also put $\ga=1$. Thus, assume 
$\ga_i\in[0,1]$ and
$\ga_1+\ga_2=1$. So, having started from fourteen, essentially we are now left with seven free parameters (think of all greek symbols of dimension $[time]\inv$ being divided by $\ga$).

\bsn
To further classify the space of admissible parameters some formalism will be needed. Put
\begin{align}
\C	&:=\{(\al_i,\ga_i,\theta_i)\in\RR^6\mid 
\al_1+\al_2>0\,\land\, \ga_1+\ga_2=1\}
\label{C}\\
\C_+	&:=\C\cap\{(\al_i, \ga_i)\in\RRN^4\}
\label{C_+}\\
\Csig	&:=\C\cap\{\theta_1\geq0\geq\theta_2\}
\label{Csig} \\
\Cph 	&:=\C_+\cap\{\theta_i+\al_i\geq 0\,,\ i=1,2\}
\label{Cph} \\
%\Cph 	&:=\C_+\cap\hCph
%\label{Cph} \\
\Cbio	&:=\Csig\cap\Cph
\label{Cbio} 
\end{align} 
Note that for $\theta_i=0$ we have $\C_+=\Cph=\Cbio$. Denoting 
\begin{equation}
\B:=\{\bbe=(\be_1,\be_2)\in\RR^2\mid \be_2<\be_1\}.
%\qquad\quad\BN:=\B\cap\{\be_2\geq 0\}.
\label{B-space}
\end{equation}
the full parameter sets are then given by $\A:=\C\times\B$ or
$\A_x:=\C_x\times\B$,  respectively. I will also use obvious notations like $\A_{\bth=\bzero}:=\A\cap\{\th_i=0\}$ and
$\A_{\bal\geq\bzero}:=\A\cap\{\al_i\geq 0\}$. 
%So $\Abio$ and $\Aph$ are 7-dimensional, $\Aone$ is 6-dimensional, $\Asirs$ and  $\Atwo$ are 5-dimensional and $\Aheth$ is 3-dimensional\footnote{
%Beware that having normalized time scale to 
%$\ga_1+\ga_2=1$ reduces the free parameter count by one.}.

{\remark In the definition of $\C$ in \eqref{C} the border case $\al_1=\al_2=0$ (i.e. absence of constant vaccination and waning immunity rates) has been excluded, see Appendix \ref{Sec_a=0} for a short discussion. For the body of this paper I will stick with the assumption 
$\al_1+\al_2>0$.}

%\bsn
%Eventually, from an epidemiological point of view this paper analyses the following three models in more detail
\bsn
Next, it is easy to check, that for $\bfa\in\Aph$ 
the {\em physical triangle} 
\begin{equation} 
\Tph:=\{(S_1,S_2,I)\in\RR_{\geq 0}^3\mid S_1+S_2+I=1\} 
\label{Tph1} 
\end{equation}
stays forward invariant under the dynamics \eqref{dot_S_N(t)}-\eqref{dot_I_N(t)}, i.e. on $\Tph$ we have $I=0\Rightarrow\dot{I}=0$ and $S_i=0\Rightarrow\dot{S}_i\geq 0$. 
Note that 
$\theta_i+\al_i\geq 0$ in \eqref{Cph} is 
sufficient but not necessary to assure this.

{\lemma\label{Lem_Tphys}
In the SSISS model \eqref{dot_S_N(t)}-\eqref{dot_I_N(t)} the physical triangle stays forward invariant for all parameters 
$(\al_i,\be_i,\ga_i,\theta_i)\in\Aph$, also including the border case 
$\al_1=\al_2=0$.
}\qed

\bsn
We are now ready to state a main result of this paper. Assuming $\nu_1=\nu_2$ the normalization procedure to be introduced in Section \ref{Sec_Repl.no} will further reduce the number of essential parameters from seven to five. This means, 
%there is a new 5-dimensional parameter space $\D$ and a surjective submersion 
%$\A\ni\bfa\mapsto\bx(\bfa)\in\D$, such that SSISS models with parameters $\bfa\in\A$ are isomorphic to normalized models with parameters $\bx(\bfa)\in\D$. Hence, 
SSISS models fall into isomorphy classes 
mapping to the same normalized system.
%$\A_\bx\subset\A$ given by the pre-images of 
%$\bx\in\D$.
It turns out, that these isomorphy classes coincide with orbits under a parameter symmetry group $G_S$ acting simultaneously on phase $\P$ and parameter space $\A$, such that
parameters for the normalized system are naturally identified as elements of $\A/G_S$.
%$\bx(\bfa\li\g)=\bx(\bfa)$ and 
%$\D\cong\A/G_S$.
%%
% As a particular result, if 
%$\theta_1\geq \theta_2$ or $\theta_1\theta_2>0$
%then $\theta_i$ may be "gauged to zero", meaning that under this condition 
%every isomorphy class contains precisely one representative in $\A_{\bth=\bzero}$. 
%To be more precise put $c:=\be_1+\be_2+\theta_1+\theta_2$ and $\ep:=\be_1\be_2+\be_1\theta_2+\be_2\theta_1$ and define 
%\begin{equation}
%\A_B:=\A\cap\{c^2>4\ep\}
%\label{A_B}
%\end{equation}
%As we will see in \Eqref{A_sig} we have 
%$\Abio\subset\Asig\subset\A_B$ and the following holds true.

{\theorem\label{Thm_gaugefix}
For $\by=(S_1,S_2,I)^T\in\RR^3$ and parameter values 
$\bfa=(\bal,\bbe,\bga,\bth)\in\A$ denote $\dot{\by}=\bF_\bfa(\by)$ the dynamical system \eqref{dot_S_N(t)}-\eqref{dot_I_N(t)} with vector field $\bF_\bfa:\RR^3\rto\RR^3$. 
Let $G_S\subset GL_+(\RR^2)$ be the subgroup acting on 
$\bS\in\RR^2$ from the left and leaving $S_1+S_2$ invariant. 
\begin{itemize}
\item[i)]
Then there exists a free right action $\li:\A\times G_S\rto\A$ such that $\A$ becomes a principal $G_S$-bundle and 
\begin{equation}
\bF_\bfa\circ\bT_\bg=\bT_\bg\circ\bF_{\bfa\li\bg}\,,\qquad
\bT_\bg:=
\begin{pmatrix}
\multicolumn{2}{c}{\multirow{2}{*}{$\bg$}}&0\\
&&0\\
0&0&1
\end{pmatrix}\,,\qquad\forall(\bfa,\bg)\in\A\times G_S.
%,\qquad\bra\one|\bT_\bg=\bra\one|
\label{T_g}
\end{equation}
\item[ii)]
Put 
$\bj:=\left(
\begin{smallmatrix}
0&1\\1&0
\end{smallmatrix}
\right)$
and for $\bg\in G_S$ denote $\bar{\bg}:=\bj\bg\bj\in G_S$. Viewing 
$\bal,\bga,\bth\in\RR^2$ as column vectors and $\bbe\in\B$ as a row vector and writing 
$\bfa\li\bg=\bfa'=(\bal',\bbe',\bga',\bth')$ we have
\begin{align*}
\bal'	&=\bar{\bg}\inv\bal,
		&\bth'&=\bar{\bg}\inv\bth+\bvth\\
\bga'	&=\bg\inv\bga,
		&\bvth&=\frac{1}{\be'_1-\be'_2}
		\begin{pmatrix}
		-(\be_1-\be_1')(\be_2-\be_1')\\
		(\be_1-\be_2')(\be_2-\be_2')
		\end{pmatrix}
\\
\bbe'&=\bbe\bg
\end{align*}
% then 
%$
%\bbe'=\bbe\bg
%\equiv(\sum_i\be_ig_{i1},\sum_i\be_ig_{i2}).
%$
\item[iii)]
The $G_S$-right action 
$\B\times G_S\ni(\bbe,\bg)\mapsto\bbe\bg\in\B$ is free and transitive and $\A\cong\A/G_S\times\B$ as trivial principal fiber bundles.
\item[iv)]
Put $\bS'=\bg\inv\bS$. Then $\bra\bbe|\bS\ket=\bra\bbe'|\bS'\ket\equiv\Reff$ and therefore 
$\dot{X}_{\mathrm{rep}}=f_\bfa(X_{\mathrm{rep}},I)$ where $f_\bfa=f_{\bfa\li\bg}$ is $G_S$-invariant, i.e. it only depends on $\A/G_S$.
\item[v)]
If $\theta_1\geq\theta_2$ or 
$\theta_1\theta_2>0$\footnote{Actually these conditions are sufficient but not necessary. For a weaker condition see Section \ref{Sec_results}.}, then there exists
$\bg\in G_S$ such that 
$\bfa':=\bfa\li\bg\in\A_{\bth=\bzero}$, i.e. the parameters $\theta_i$ may be ``gauged to zero''. If in this case 
$\bfa\in\Abio$ then also $\bfa'\in\Abio$.
\end{itemize}
}

{\remark\label{Rem_T_g}
As we will see, although the linear transformation $\T_\bg$ preserves the condition $S_1+S_2+I=1$, it does not necessarily leave 
$\RRN^3$ (and hence $\Tph$) invariant.}

{\remark\label{Rem_f_a}
Since $\dim G_S =2$ we have $\dim \A/\G_S =\dim \A -2$. So, using $(\Reff,I)$ as independent coordinates in $\Tph$, the number of essential parameters of the SSISS dynamical system reduces from seven to five.}

\bsn
Parts i)-iv) of Theorem \ref{Thm_gaugefix} will be proven in 
Corollary \ref{Cor_GS-bundle} and Lemma \ref{Lem_transformation} and part v) in Lemma \ref{Lem_A_B}. Before coming to this let me close this Section 
\begin{itemize}
\item[-]
in Subsection \ref{Sec_examples} with shortly revisiting some bench-marking models in the literature within the present framework,
\item[-]
in Subsection \ref{Sec_periodic-sol} with proving absence of periodic solutions by optimizing the methods of \autocite{BusDries90}.
\end{itemize}

\subsection{Examples from the literature\label{Sec_examples}}
For simplicity, in this subsection let me assume a compartment independent birth rate $\delta$. Formulating the dynamics for fractional variables 
$\by=(S_1,S_2,I)$ there always remains an ambiguity by adding a vectorfield vanishing on $\Tph$. In Eqs. \eqref{dot_S_N(t)}-\eqref{dot_I_N(t)} the vector field $\bF\equiv\bF_\bfa$ has the special form 
\begin{equation}
\bF(\by)=\bM\by+\bGa(\by\otimes\by),\qquad\bra\one|\bM=\bra\one|\bGa=\bzero,
\end{equation} 
where $\bM\in\RR^{3\times 3}$, $\one=(1,1,1)$ and 
$\bGa\in\Hom(\RR^3\otimes\RR^3,\RR^3)$. As is shown in Appendix \ref{Sec_vital-dynamics}, n-compartment models with at most quadratic terms and  population size varying only due to constant per capita birth and death rates may always be normalized in this way.
Using different conventions bears the risk of overlooking redundancies in parameter space. Moreover, it also makes it tedious to pin down the differences between (or equivalence of) 
various models in the literature.
Table \ref{Tab_examples} shows how the examples quoted at the beginning of this Section\footnote{
${\mathrm{Heth}}=$ \autocite{Hethcote1974, Hethcote1976, Hethcote1989};
${\mathrm{SIRI}}=$ \autocite{DerrickDriessche};
${\mathrm{BuDr}}=$ \autocite{BusDries90};
${\mathrm{SI(R)S}}=$ 10-parameter mixed SIRS/SIS model with constant population size and $\theta_2=\be_2=0$;
${\mathrm{HaCa}}=$ core system in \autocite{Had_Cast};
${\mathrm{KZVH}}=$ \autocite{KribsVel};
${\mathrm{LM}}=$ \autocite{LiMa2002};
${\mathrm{AABH}}=$ \autocite{AvramAdenane2022}.
${\mathrm{SIRI}}$ and ${\mathrm{AABH}}$ come in two versions, the subscript $1$ refers to 
$\be_S>\be_R$ and $2$ to $\be_S<\be_R$.
}
compare with each other when mapped to the present set of parameters.

\begin{table}[htpb!]
\addtocounter{footnote}{-1}
\caption{Mapping models in the literature\protect\footnotemark\ 
expressed in non-normalized variables 
$(\mathbb{S}_1, \mathbb{S}_2, \mathbb{I})$
to the present choice of parameters.  The column $\#$ counts the number of free parameters in the original models.
After passing to fractional variables $(S_1,S_2,I)$  and tilde parameters, \Eqref{notation} or \Eqref{notation2},
and resetting time scale to $\tilde{\ga}=1$,
the column $\#_{\mathrm{eff}}$ counts the number of effectively independent parameters as determined in Eqs. \eqref{Heth}-\eqref{AABH}.
}
\label{Tab_examples} 
$$
\begin{array}{l|cccccccccccc|cc}
&\al_1&\al_2&\be_1&\be_2&\ga_1&\ga_2&
\delta&\mu_1&\mu_2&\mu_I&q_1&q_2&\#&\#_{\mathrm{eff}}\\
\hline
\mathrm{Heth}&0&0&\ok&0&0&\ok&
\multicolumn{4}{c}{\delta=\mu_1=\mu_2=\mu_I}&1&0&3&2
\\\hline
\mathrm{SIRI_1}&0&0&\ok&\ok&0&\ok&\ok&\multicolumn{2}{c}{\mu_1=\mu_2}&\ok&1&0&6&3
\\\hline
\mathrm{SIRI_2}&0&0&\ok&\ok&\ok&0&\ok&\multicolumn{2}{c}{\mu_1=\mu_2}&\ok&0&1&6&3
\\\hline
\mathrm{BuDr}&0&\ok&\ok&0&0&\ok&\ok& \ok&\ok&\ok&1&0&7&5
\\\hline
\mathrm{SI(R)S}&\ok&\ok&\ok&0&\ok&\ok&
\multicolumn{4}{c}{\delta=\mu_1=\mu_2=\mu_I}&\ok&\ok&7&4
\\\hline
\mathrm{HaCa}&\ok&0&\ok&\ok&\ok&\ok&
\multicolumn{4}{c}{\delta=\mu_1=\mu_2=\mu_I}&1&0&6&5
\\\hline
\mathrm{KZVH}&\ok&\ok&\ok&\ok&\ok&\ok&
\multicolumn{4}{c}{\delta=\mu_1=\mu_2=\mu_I}&1&0&7&5
\\\hline
\mathrm{LM}&\ok&\ok&\ok&0&\ok&0& \ok&\multicolumn{2}{c}{\mu_i=f(N)}&\ok&\ok&\ok&8&5
\\\hline
\mathrm{AABH_1}&\ok&\ok&\ok&\ok&0&\ok&\ok&\multicolumn{2}{c}{\mu_1=\mu_2\footnotemark}&\ok&1&0&8&5
\addtocounter{footnote}{-1}
\\\hline
\mathrm{AABH_2}&\ok&\ok&\ok&\ok&\ok&0&\ok&\multicolumn{2}{c}{\mu_1=\mu_2\footnotemark}&\ok&0&1&8&5
\\\hline
\end{array}
$$
\end{table}

\footnotetext{\label{FN_AABH} The bulk of results in Section 5 and 6 of \autocite{AvramAdenane2022} assumes $\mu_1=\mu_2$.}

\bsn
Applying the transformations \eqref{notation} or \eqref{notation2}, respectively, maps the above 11-parameter set 
to the redundancy-free 6-parameter set
$(\tilde{\al}_i,\tilde{\beta}_i,\tilde{\ga}_i)$. After resetting time scale to $\tilde{\ga}\equiv\tilde{\ga}_1+\tilde{\ga}_2=1$ the classification of the above models looks as follows:
\begin{align}
\Aheth	&=\Abio\cap\A_{\bth=\bzero}
\cap\{\tilde{\al}_1=0\,\land\,\tilde{\ga}_2>0\,\land\,
\tilde{\ga}_1=\tilde{\al}_2\,\land\,\tilde{\beta}_2=0\}
%\\
%&=\A_{\mathrm{SIRI_1}}\cap\{\tilde{\beta}_2=0\}
\label{Heth}
\\
\A_{\mathrm{SIRI}_i} &=\Abio\cap\A_{\bth=\bzero}
\cap\{\tilde{\al}_i\,=0\,\land\,\tilde{\ga}_j>0\,\land\,
\tilde{\ga}_i\,=\tilde{\al}_j,\,j\neq i\}
\label{SIRI}\\
\A_{\mathrm{BuDr}} &=\Abio\cap\A_{\bth=\bzero}
\cap\{\tilde{\al}_1=0\,\land\,\tilde{\ga}_2>0
\,\land\,\tilde{\beta}_2<0\}\footnotemark
\label{BuDr}\\
\Asirs	&=\Abio\cap\A_{\theta_2=0}\cap\{\tilde{\beta}_2=0\}
\label{SIRS} \\
\A_{\mathrm{KZVH}} &=\Abio\cap\A_{\bth=\bzero}
\cap\{\tilde{\be}_2>0\}=\A_{\mathrm{HaCa}}
\label{KZVH}\\
\A_{\mathrm{LM}} &=\Abio\cap\A_{\bth=\bzero}
\cap\{\tilde{\be}_2<0\,\land\,\tilde{\ga}_1>0\}
\label{LM}\\
\A_{\mathrm{AABH}_i} &=\Abio\cap\A_{\bth=\bzero}\cap
\{\tilde{\ga}_j>0,\,j\neq i\}
\label{AABH}
%\Aone	&=\Abio\cap\{\be_2>0,\theta_2=0\}
%\label{A_one} \\
%\Atwo 	&=\Abio\cap\{\be_2>0,\al_1=0, \al_2=-\theta_2\}
%\label{A_two} 
\end{align} 
\footnotetext{To be comparable \Eqref{BuDr} refers to the sub-case $\mu_1=\mu_2$ in \autocite{BusDries90}, so
$\dim{\A_{\mathrm{BuDr}}}=4$. Allowing also an excess mortality 
$\mu_2-\mu_1>0$ gives 
$\#_{\mathrm{eff}}=5$ in Table \ref{Tab_examples}.
\label{FN_BuDr}}

\noindent
The dimensions of these parameter spaces are displayed in the last column of Table \ref{Tab_examples}\addtocounter{footnote}{-1}\footnotemark. 
To verify Eqs. \eqref{Heth}-\eqref{AABH} the following explanations should suffice.
\begin{itemize}

\item The SIRI model of \autocite{DerrickDriessche} with varying population requires 
$\al_i=\ga_1=0$. Since for $\be_R>\be_S$ the mapping to the SISS model permutes $1\lra 2$ (i.e. maps $R\rto S_1$ and $S\rto S_2$), if $\be_R<\be_S$ we get 
$\tilde{\al}_1=0$, $\tilde{\al}_2=\tilde{\ga}_1=\delta$ and
$\tilde{\ga}_2=\ga_2>0$, and if $\be_R>\be_S$ we get
$\tilde{\al}_2=0$, $\tilde{\al}_1=\tilde{\ga}_2=\delta$ and
$\tilde{\ga}_1=\ga_1>0$.

\item
The SIRS model of \autocite{BusDries90} differs from SIRI by allowing 
$\al_2>0$ and $\mu_1<\mu_2$, but in turn it requires 
$\be_S>\be_R=0$. Thus, we have $\tilde{\al}_1=0$ and 
$\tilde{\ga}_1=\delta$ as in SIRI$_1$, but 
$\tilde{\al}_2=\al_2+\delta$ becomes independent. If, for comparison, we restrict to $\mu_1=\mu_2=\mu$ then $\be_2=0$ implies 
$\tilde{\be}_2=-\Del\mu_I\leq 0$. 
%Thus the SIRI model of \autocite{DerrickDriessche} appears as a true sub-case of the SIRS model of \autocite{BusDries90}. Note that this may come as a surprise, since \autocite{DerrickDriessche} had published their paper three years later and also argue themselves, that it should be more complicated than \autocite{BusDries90}. Also note that if in BuDr we also put $\mu_1=\mu_2$, then 
%$\A_{\mathrm{BuDr}}$ restricts to $\tilde{\be}_2<0$, whereas 
%$\A_{\mathrm{SIRI}}$ doesn't.

\item
If $q_1>0$ then one of the three parameters 
$(\ga_1,\al_2,\delta)$ always becomes redundant. So the models of
\autocite{Had_Cast} and \autocite{KribsVel} are isomorphic, in spite of the latter containing the additional immunity waning rate $\al_2$. Also, they both satisfy $\tilde{\be}_2=\be_2>0$.

\item 
Putting $q_2=1$ in the SIS-type model of \autocite{LiMa2002} the mapping $(\al_1,\al_2,\ga_1,\delta)\mapsto (\tilde{\al}_i, \tilde{\ga}_i)$ is bijective. Also, the authors have defined $\mu_i=f(N)$ and $\mu_I=f(N)+\Del\mu_I$. Hence, the only restrictions in this model are 
$\tilde{\be}_2=-\Del\mu_I<0$ and $\tilde{\ga}_1>0$.
\end{itemize}

\noindent
In summary we get the following conclusions, which apparently have not yet been realized in the literature.

\begin{corollary}\label{Cor_Avr=JiLi+KrZa} 
Assume $\mu_1=\mu_2=:\mu$ and put $\Del\mu_I:=\mu_I-\mu$.
\begin{itemize}
\item[i)]For $\be_1>\be_2=\Del\mu_I$ the SIRI model of 
\autocite{DerrickDriessche} is isomorphic to Hethcote's classic endemic model.
\end{itemize}
\noindent
Moreover, restricting to $\tilde{\ga}_1>0$ and 
$\be_2\neq \Del\mu_I$ we have 
\begin{itemize}
\item[ii)]
The SIRS-type model of \autocite{BusDries90} reduces to a sub-case of the SIS-type model of \autocite{LiMa2002}, which in turn covers Sectors \RN{3}-\RN{7} of the SSISS model at $\theta_i=0$.
\item[iii)]
The models of \autocite{Had_Cast} and \autocite{KribsVel} are isomorphic and cover Sector \RN{1} of the SSISS model at 
$\theta_i=0$.
\item[iv)]
The models of \autocite{LiMa2002} and 
\autocite{Had_Cast, KribsVel} only differ by the sign of 
$\tilde{\be}_2$. 
\item[v)]
Their disjoint union covers the SIRI model of \autocite{DerrickDriessche} and coincides with the model of \autocite{AvramAdenane2022}.
\end{itemize}
\end{corollary}
An equivalent formulation of Corollary \ref{Cor_Avr=JiLi+KrZa} based on normalized parameters and variables is given in 
Corollary \ref{Cor_Avr=JiLi+KrZa_2} in Section \ref{Sec_examples2}.

\subsection{Absence of periodic solutions\label{Sec_periodic-sol}}
In this subsection I will specify parameter ranges guaranteeing absence of periodic solutions by optimizing methods from \autocite{BusDries90} (see also \autocite{BusDries91, DerrickDriessche}) for the present situation, including 
$\theta_i\neq 0$. To start with, the Busenberg-Driessche version of the classical Bendixson–Dulac Theorem may be given the following alternative formulation

\begin{lemma}\label{Lem_BusDries90} \autocite{BusDries90}
Let $\bF:\RR^3\rto\RR^3$ be smooth in a neighborhood of $\Tph$ and assume $\Tph$ forward invariant under the flow of $\dot{\by}=\bF(\by)$. 
Assume there exists a smooth scalar function $u(\by)$ defined in a neighborhood of $\Tph$ such that 
\begin{equation}
\Psi(\by):=\nabla\cdot(u\bF)(\by)
-(\by\cdot\nabla)(u\sum_iF_i)(\by)\leq 0\,,\qquad\forall\by\in\Tph
\label{Psi}
\end{equation}
and $\Psi(\by)<0$ for some $\by\in\Tph$. Then in 
$\Tph\setminus\partial\Tph$ there exist no periodic solutions, homoclinic loops or oriented phase polygons of the dynamical system 
$\dot{\by}=\bF(\by)$.
\end{lemma} 
\begin{proof}
Put $\mathbf{1}:=(1,1,1)$ and $\bg:=\by\times u\bF$. 
Then $\bg\cdot\bF=0$ and 
$\bra\mathbf{1}\mid\nabla\times\bg\ket|_{\Tph}=\Psi|_{\Tph}$, where the second identity easily follows from 
$\bra\mathbf{1}\mid\bF\ket|_{\Tph}=0$. Now the claim follows by Stoke's Theorem as in the proof of Theorem 4.1 of \autocite{BusDries90}.
\end{proof}

{\remark\label{Rem_tildeF}
In Lemma \ref{Lem_tildeF} in Appendix \ref{Sec_vital-dynamics} it is shown that for models with constant per capita birth and death rates one may always replace $\bF$ by $\tilde{\bF}$ obeying 
$\bF|_{\Tph}=\tilde{\bF}|_{\Tph}$ and $\bra\one\mid\tilde{\bF}\ket=0$ also outside $\Tph$. So in this case the second term in \eqref{Psi} vanishes and the condition 
$\nabla(u\tilde{\bF})\leq 0$ looks like in the classical Bendixson-Dulac theorem.
}

\bsn
As in \autocite{BusDries90} putting $\by=(S_1,S_2,I)$ and 
$u=1/(S_1S_2I)$ we now apply this to the dynamical system Eqs. \eqref{dot_S_N(t)}-\eqref{Replacement-Nr_N(t)}. We have 
$u\bF(\by)=u\bM\by+u\bff(\by)$ where
\begin{equation}
\bM = 
\begin{pmatrix}
-\tilde{\al}_1&\tilde{\al}_2&\tilde{\ga}_1\\
\tilde{\al}_1&-\tilde{\al}_2&\tilde{\ga}_2\\
0&0&-1
\end{pmatrix},
\qquad
(u\bff)(\by) =
\begin{pmatrix}
-(\tilde{\be_1}+\theta_1)/S_2+\theta_2/S_1+(\nu_1-\nu_2)/I\\
-(\tilde{\be_2}+\theta_2)/S_1+\theta_1/S_2+(\nu_2-\nu_1)/I\\
\tilde{\be}_1/S_2+\tilde{\be}_2/S_1
\end{pmatrix}.
\end{equation}
Here the time scale normalization 
$\tilde{\ga}_1+\tilde{\ga}_2=1$ is understood.
\begin{theorem}
\label{Thm_periodic}
Under the following conditions there exist no periodic solutions, homoclinic loops or oriented phase polygons of the SSISS system \eqref{dot_S_N(t)}-\eqref{Replacement-Nr_N(t)} in $\Tph$.
\begin{itemize}
\item[i)]
$(\tilde{\al}_i,\tilde{\ga}_i,\theta_i)\in\RRN^6$.
\item[ii)]
$(\tilde{\al}_i,\tilde{\ga}_i,\theta_i)\in\Cbio$ and
$\nu_1=\nu_2$.
\end{itemize}
\end{theorem}
\begin{proof}
First note that
$\tilde{\ga}_1+\tilde{\ga}_2=1$ implies that the boundary lines 
$\{S_1=0\}$ and $\{S_2=0\}$ cannot both be forward invariant.
Hence, $\partial\Tph$ cannot be a phase polygon. 
Next, the second term in \eqref{Psi} vanishes, because we have 
$\bra\mathbf{1}\mid\bF\ket=0$ also outside of $\Tph$. We are left to compute 
$\nabla\cdot(u(\by)\bM\by)=-\sum_{i\neq j}M_{i,j}y_j/y_i<0$ and 
$\nabla\cdot\bff=-\theta_2/S_1^2-\theta_1/S_2^2$.  Part i) follows by Lemma \ref{Lem_Tphys} and Lemma \ref{Lem_BusDries90}. 
The proof of part ii) relies on the normalization formalism of Section \ref{Sec_Repl.no} and follows from Corollary \ref{Cor_periodic}. 
\end{proof}

\begin{remark}
Note that Theorem \ref{Thm_periodic}ii) doesn't follow directly from  Theorem \ref{Thm_gaugefix}, because there the equivalence transformation $\bT_\bg$ need not preserve $\Tph$, see also Remark \ref{Rem_T_g}.
\end{remark}

\begin{remark}
Usually in the literature on models with constant per capita birth and death rates the vector field $\bF$ appears in the form
$\bF=\bF_M+\bff$, where 
$\bF_M=\bM\by-\bra\mathbf{1}\mid\bM\by\ket\by$, the second term being nonzero. This makes computations more involved but still yields $\Psi_M|_{\Tph}\equiv\nabla\cdot(u\bF_M)|_{\Tph}
-(\by\cdot\nabla)\bra\mathbf{1}\mid u\bF_M\ket|_{\Tph}=
-\sum_{i\neq j}M_{i,j}y_j/y_i$, see Eq. (3.8) in 
\autocite{DerrickDriessche}. The fact that  
$\bM$ may be chosen to satisfy $\bra\mathbf{1}|\bM=\bzero$ 
(Lemma \ref{Lem_tildeF} in Appendix \ref{Sec_vital-dynamics}, see also remark \ref{Rem_tildeF}) is rarely noticed in the literature.
\end{remark}

\section{Normalization\label{Sec_Repl.no} }
\subsection{Phase space\label{Sec_Phase-Space}}
From now on we drop again the tilde above parameters and also require $\nu_1=\nu_2$. To proceed one has to choose suitable coordinates 
$(X,Y)$ on a phase space $\P\supset\Tph$. Let's first do some linear algebra. Put $V=\RR^2$ and consider
$\bS\equiv|\bS\ket=\left(
	\begin{smallmatrix}
	\S_1\\ \S_2
	\end{smallmatrix}
	\right)
$,
$\bal\equiv|\bal\ket=\left(
	\begin{smallmatrix}
	\al_1\\ \al_2
	\end{smallmatrix}
	\right)
$,
$\bga\equiv|\bga\ket=\left(
	\begin{smallmatrix}
	\ga_1\\ \ga_2
	\end{smallmatrix}
	\right)
$,
$\bth\equiv|\bth\ket=\left(
	\begin{smallmatrix}
	\th_1\\ \th_2
	\end{smallmatrix}
	\right)
$
as elements of $V$ (``ket-'' or ``column-'' vectors). Denote 
\begin{equation}
\bfe\equiv\bra\bfe|\ :=(1,1)\,,\qquad
\bbe\equiv\bra\bbe|\ :=(\be_1,\be_2)
\label{bra-basis}
\end{equation}
as a basis in the dual space $V^*$ (``bra-'' or ``row-'' vectors). 
Putting $\bL(\bbe,\bth):=\bD(\bbe)+\bE(\bth)$ we then have
\begin{equation}
\bra\bfe|\bE(\al)=0,\quad\bra\bfe|\bL(\bbe,\bth)=\bra\bbe|,\quad
\bra\bfe\mid\bga\ket=1
%\quad\bra\bbe|\bS\ket=1-I,\quad
%\bra\bbe\mid\bS\ket=\Reff
\end{equation}
where $\bra\cdot\mid\cdot\ket$ denotes the dual pairing $V^*\otimes V\rto\RR$. Generalizing this setting, pick 
$(\bfe,\bbe)$ any oriented\footnote{The requirement of being oriented (with respect to a given orientation in $V$) is a coordinate free version of the condition 
$\be_2<\be_1$.}
basis in $V^*$ and $\bga\in V$ satisfying 
$\bra\bfe\mid\bga\ket=1$. Denote
$\E\subset\End V$ the right ideal anihilated by  $\bra\bfe|$ and
$\L:=\{L\in\End V\mid \bra\bfe|\bL=\bra\bbe|\}$. On 
$V\times\RR=\RR^3$ consider the dynamical system
\begin{align}
\dot{\bS}&=-\left[\bE+I\bL\right]\bS+I\bga,\qquad 
\bS\in V,\,\bE\in\E,\,\bL\in\L,\,
\label{dot_S_coord-free}\\
\dot{I}&=(X-1)I,\qquad\qquad\qquad I\in\RR,\,
X:=\bra\bbe\mid\bS\ket.
\label{dot_I_coord-free}
\end{align}
Fixing $\bfe$ and varying $(\bbe,\bE,\bL, \bga)$ under the above constraints defines a $7$-parameter dynamical system which in fact provides a coordinate free reformulation of the SSISS model \eqref{dot_S}. Note that the conditions imply 
$\bra\bfe\mid\dot{\bS}\ket+\dot{I}\equiv
\dot{S}_1+\dot{S}_2+\dot{I}=0$, so the dynamics \eqref{dot_S_coord-free}-\eqref{dot_I_coord-free} leaves the cosets 
$\{\bra\bfe\mid\bS\ket+I=\const\}\subset\RR^3$ invariant. Since $I=0$ implies $\dot{I}=0$ also the half spaces $\{I\in\RR_\pm\}$ as well as the plane $\{I=0\}$ stay invariant. 

{\definition\label{Def_extended_SSISS}
The dynamical system \eqref{dot_S_coord-free}-\eqref{dot_I_coord-free} on phase space $\P=\{(\bS,I)\in V\times\RRN\mid \bra\bfe\mid\bS\ket+I=1\}$ with parameter space $\A=\C\times\B$ is called the {\em extended SSISS model}.
}
{\remark
The extension to negative values of variables $S_i$ and parameters $\bfa$ is needed to construct the symmetry operation of $G_S$ in Theorem \ref{Thm_gaugefix}.
}
%Defining 
%$$\bar{\P}:=\{(\bS,I)\in V\times\RR_{\leq 0}\mid \bra\bfe\mid\bS\ket+I=1\}$$
% we will also have a reflection symmetry 
%$\P\rto\bar{\P}$ leaving the extended SSISS dynamics invariant provided parameters are also reflected appropriately, 
%$\A\ni\bfa\mapsto\bar{\bfa}\in\A$, see the next subsection and in more detail also part \RN{2} of this work \autocite{Nill3}.

\subsection{Canonical coordinates\label{Sec_Can-Coord}}
Putting $I:=1-\bra\bfe\mid\bS\ket$ and using $\bS$ as independent coordinates on $\P$ \Eqref{dot_I_coord-free} becomes redundant and we end up with a two-dimensional system. However, based on the coordinate free formulation \eqref{dot_S_coord-free}-\eqref{dot_I_coord-free}, there is another natural set of 
{\em canonical coordinates} for this system. Put
\begin{equation}
X:=\bra\bbe\mid\bS\ket,\qquad Y:=\bra\bfe\mid\bS\ket,
\label{can-coord}
\end{equation}
or equivalently choose the basis dual to \eqref{bra-basis} in $V$ 
\begin{equation}
\bfe^\perp\equiv|\bfe^\perp\ket\ :=\frac{1}{\be_1-\be_2}
\begin{pmatrix}
1\\ -1
\end{pmatrix}\,,
\qquad
\bbe^\perp\equiv|\bbe^\perp\ket\ :=\frac{1}{\be_1-\be_2}
\begin{pmatrix}
-\be_2\\ \be_1
\end{pmatrix}
\label{ket_basis}
\end{equation}
Hence we have $X\equiv\Reff$, $Y\equiv S_1+S_2$ and 
\begin{equation}
\bS=X\bfe^\perp+Y\bbe^\perp.
\label{S_from_X}
\end{equation}

{\lemma\label{Lem_Xrep-ODE}
In canonical coordinates the extended SSISS model becomes 
\begin{eqnarray}
\dot{X} &=& (-aX +b)+(-cX+d)I-\ep I^2\,,\label{dot_X}
\\
\dot{Y} &=& (1-X)I=-\dot{I}\,,\label{dot_Y} 
\end{eqnarray}
where $I=1-Y$
%The functions $A_i$ are of the form
%\begin{equation}
%A_0(X) = -aX +b\,,\qquad
%A_1(X) = -cX+d\,.
%\label{f} 
%\end{equation}
and where the new parameters are given by
\begin{align}
a	&:=\al_1+\al_2\label{a} \\
b	&:=\al_2\beta_1+\al_1\beta_2\label{b} \\
c	&:=\beta_1+\beta_2+\theta_1+\theta_2\label{c} \\
d	&:=\ga_1\be_1+\ga_2\be_2 -b+\ep\label{d} \\
\ep&:=\beta_1\beta_2+\beta_1\theta_2+\beta_2\theta_1
\label{ep}\,.
\end{align} 
}
\begin{proof}
In canonical coordinates 
the matrices $\bE(\bal)$ and $\bL(\bbe,\bth):=\bD(\bbe)+\bE(\bth)$ take the normal form 
\begin{equation}
\bE(\bal)=
\begin{pmatrix}
a & -b \\ 0 & 0
\end{pmatrix}\,,
\qquad
\bL(\bbe,\bth)=
\begin{pmatrix}
c & -\ep \\ 1 & 0
\end{pmatrix}.
\label{normal_matrix}
\end{equation}
Using 
$|\bga\ket=
(\be_1\ga_1+\be_2\ga_2)|\bfe^\perp\ket+|\bbe^\perp\ket$ the claim follows by straightforward calculation.
\end{proof}
\bsn
The canonical form of the SSISS dynamical system \eqref{dot_X}-\eqref{dot_Y} will also be called the {\em RN-dynamical system}
(RN = replacement number). Beware that unless $\be_2\geq 0$ the ``would-be'' replacement number  $X$ may take negative values even for 
$S_i\geq 0$. In fact, in canonical coordinates the physical triangle takes the form
\begin{align} 
\Tph(\bbe)&=\{(X,Y)\in\RR\times[0,\,1]\mid 
\beta_2Y\leq X\leq\beta_1Y\}\,\notag\\
&=\{(X,I)\in\RR\times[0,\,1]\mid 
\beta_2(1-I)\leq X\leq\beta_1(1-I)\}.
\label{Tph2} 
\end{align}
So in $(X,I)$-space $\Tph$ is given by the corners 
$\bT_<=(\be_2,0)$, $\bT_>=(\be_1,0)$ and $\bT_\wedge=(0,1)$. To stay with epidemiological conventions, from now on I will use $X\equiv\Reff$ and $I\equiv 1-Y$ as independent variables, in terms of which  phase space is now given by 
$$\P=\{(X,I)\in\RR\times\RRN\}.
$$
%This allows to specify the before mentioned reflection symmetry. We have 
%$\bar{\P}=\RR\times\RR_{\leq 0}$ and for $\bx=(a,b,c,d,\ep)$ denote 
%\begin{equation}
%\bar{\bx}:=(a,b,-c,-d,\ep).\label{xbar}
%\end{equation} 
%Then obviously 
%$(X,I)\in\P$ solves the ODE \eqref{dot_X}-\eqref{dot_Y} at parameter values $\bx$ if and only if $(X,-I)\in\bar{\P}$ solves the same ODE at parameter values $\bar{\bx}$. The fact that 
%$\bx\mapsto\bar{\bx}$ provides an involution on the image of the transformation \eqref{a}-\eqref{ep}, whence also on $\A$, follows from Lemma \ref{Lem_new-par} below.
Also note that in canonical coordinates the 
dynamics is reduced from seven to five parameters, i.e. the system  no longer depends on $\bbe$. So, the role of 
$\bbe$ is reduced to fixing the image of physical triangles
$\Tph$ in canonical coordinates.
Equivalently this means that fixing $\bx=(a,b,c,d,\ep)$ and varying $\bbe\in\B$ we get an equivalence class of isomorphic dynamical systems, albeit physical triangles are not mapped onto each other under these isomorphisms. 

\begin{proposition}\label{Prop_SS'}
For $\bfa,\bfa'\in\A$, $\bfa=(\bal,\bbe,\bga,\bth)$ and
$\bfa'=(\bal',\bbe',\bga',\bth')$, assume  
$\bx(\bfa)=\bx(\bfa')$. Following \Eqref{S_from_X} put 
\begin{equation}
\bS:=X\bfe^\perp(\bbe)+(1-I)\bbe^\perp\,,\qquad
\bS':=X\bfe^\perp(\bbe')+(1-I)\bbe'^\perp\,.
\label{SS'}
\end{equation}
Then $S_1+S_2=S'_1+S'_2=1-I$ and $\bS=\bg\bS'$ where 
$\bg\in GL_+(\RR^2)$ is uniquely defined by
\begin{align}
\bg&=|\bbe^\perp\ket\bra\bfe|+|\bfe^\perp(\bbe)\ket\bra\bbe'|=
\frac{1}{\be_1-\be_2}
\begin{pmatrix}
\be'_1-\be_2 & \be'_2-\be_2\\
\be_1-\be'_1 & \be_1-\be'_2\,,
\end{pmatrix}
\label{g}
\end{align}
implying $\det\bg=(\be'_1-\be'_2)/(\be_1-\be_2)>0$. Moreover, 
$(\bS,I)$ satisfies the SSISS dynamics \eqref{dot_S_coord-free}-\eqref{dot_I_coord-free} at parameter values $\bfa$ iff $(\bS',I)$ satisfies it at parameter values 
$\bfa'$.
\end{proposition}

\begin{proof}
\Eqref{SS'} implies $\bra\bfe|\bS\ket=\bra\bfe|\bS'\ket=1-I$ and
$\bra\bbe|\bS\ket=\bra\bbe'|\bS'\ket=X$. Hence, $\bg$ must satisfy $\bra\bfe|\bg=\bra\bfe|$ and $\bra\bbe|\bg=\bra\bbe'|$ with unique solution \eqref{g}.
\end{proof}

\begin{remark}\label{Rem_GS-action}
Apparently we have $\bg\in G_S:=\{\bg\in GL_+(\RR^2)\mid\bra\bfe|\bg=\bra\bfe|\}$ and by
\Eqref{g}  $\bbe\mapsto\bbe\bg$ defines a transitive and free  right action of $G_S$ on $\B$\footnote{Note that $\dim G_S=2$. The parametrization of $\bg$ in \eqref{g} is redundant by invariance under $(\be_1,\be_2)\mapsto(\be_1+\lambda,\be_2+\lambda)$ and 
$(\be_1,\be_2)\mapsto(\chi\be_1,\chi\be_2)$, 
$(\lambda,\chi)\in\RR\times\RR_+$.}.
In Corollary \ref{Cor_GS-bundle} below this action will be transported to a free 
$G_S$-action on $\A$, thus proving parts i)-iv) of Theorem \ref{Thm_gaugefix}.
\end{remark}

\subsection{Main results\label{Sec_results}}
In this subsection we study the constraints on the new parameters $\bx:=(a,b,c,d,\ep)$ and admissible ranges of 
$\bbe$ - or equivalently $\Tph(\bbe)$ - for given values of $\bx$, which will finally lead to a proof of Theorems \ref{Thm_gaugefix} and \ref{Thm_periodic}. Recalling $\A\equiv\C\times\B$ denote 
\begin{equation}
\phi:\A\ni\bfa\mapsto (\bx(\bfa),\bbe)
\in\D\times\B,\qquad \D:=\RR_+\times\RR^4
\label{phi}
\end{equation}
where $\bx(\bfa)$ is given by \eqref{a}-\eqref{ep}. The proof of the following Lemma is by straight forward calculation and hence omitted.

{\lemma\label{Lem_new-par}
The map $\phi:\A\rto\D\times\B$ provides a diffeomorphism with   
$\phi\inv$ given by
\begin{equation}
\al_i =\frac{b-a\beta_i}{\beta_j-\beta_i}\,,\qquad 
\gamma_i =\frac{d+b-\ep-\beta_j}{\beta_i-\beta_j}\,,\qquad
\theta_i =\frac{\beta_i^2 -c\beta_i +\ep}{\beta_j-\beta_i}\,,\qquad j\neq i\label{inv_param} 
\end{equation}
}
\qed

\begin{corollary}\label{Cor_GS-bundle}
Consider $\D\times\B$ as a trivial principal $G_S$-bundle with fiber $\B$ and $G_S$ right action 
$(\bx,\bbe)\li\bg:=(\bx,\bbe\bg)$, see Remark \ref{Rem_GS-action}. Defining $\bfa\li\bg:=\phi\inv(\bx(\bfa),\bbe\bg)$ we get  an isomorphic $G_S$-bundle structure on $\A$.
Putting $\by:=(S_1,S_2,I)$ 
 and writing the dynamical system \eqref{dot_S_coord-free}-\eqref{dot_I_coord-free} with parameters $\bfa\in\A$ as $\dot{\by} =\bF_{\bfa}(\by)$, Proposition \ref{Prop_SS'} becomes 
$$
\bF_\bfa\circ\bT_\bg=\bT_\bg\circ\bF_{\bfa\li\bg}\,,\qquad
\bT_\bg:=\bg\oplus\id
%\begin{pmatrix}
%\multicolumn{2}{c}{\multirow{2}{*}{$\bg$}}&0\\
%&&0\\
%0&0&1
%\end{pmatrix}
\,,\qquad\bg\in G_S.
$$
This proves parts i), iii) and iv) of Theorem \ref{Thm_gaugefix}. 
\qed
\end{corollary}
\noindent
The remaining transformation rules in part ii) of 
Theorem \ref{Thm_gaugefix} now boil down to an exercise in linear algebra.
\begin{lemma}\label{Lem_transformation}
Let $\bD(\bbe)$ and $\bE(\bal)$ be given as in \Eqref{matrix} and
$\bvth(\bbe,\bbe')$ as in part ii) of 
Theorem \ref{Thm_gaugefix}. Then for all $\bg\in G_S$, $\bal\in\RR^2$ and $\bbe'=\bbe\bg\in\B$
\begin{align*}
\bE(\bar{\bg}\bal)\bg &=\bg\bE(\bal),\\
\bD(\bbe)\bg &=\bg\left[
		\bD(\bbe')+\bE(\bvth(\bbe,\bbe'))\right]
\end{align*}
\end{lemma}
\noindent
Applying these identities to the dynamical system 
\eqref{dot_S_coord-free}-\eqref{dot_I_coord-free} proves Theorem \ref{Thm_gaugefix}ii).
\qed

\begin{remark}\label{Rem_phys-Tr}
Beware that the transformation matrix $\bg$ preserves $S_1+S_2$ but not necessarily $\RRN^2$. Also, if $\bfa\in\Aph$ (or $\Abio$) and $\bx(\bfa)=\bx(\bfa')$ then it depends on $\bbe'$ whether 
$\bfa'\in\Aph$ (or $\Abio$), see Proposition \ref{Prop_Aph+Abio} below.
Hence, the above equivalencies may produce scenarios where 
$\bfa\in\Aph$ and 
$\bfa'=\bfa\li\bg\not\in\Aph$ and $\bT_\bg\inv\Tph\not\in\RRN^3$ but still $\bT_\bg\inv\Tph$ is forward invariant under the flow of 
$\bF_{\bfa'}$.
\end{remark}

\bsn
Next, on $\D$ define the functions
\begin{align}
R_0(\bx)	&:=b/a\ \qquad
\equiv\frac{\al_2\be_1+\al_1\be_2}{\al_1+\al_2},\label{R_0} \\ 
R_1(\bx)	&:=d+b-\ep
\equiv \ga_1\be_1+\ga_2\be_2\,.\label{R_1}
\end{align}
Obviously we may also use 
$\bx\equiv(a,R_0,R_1,c,\ep)\in\RR_+\times\RR^4$ as independent parameters in $\D$. 
Moreover we clearly have 
\begin{equation}
\phi(\A_+)=\{(\bx,\bbe)\in\D\times\B\mid
\be_2\leq R_i\leq \be_1\,,\,i=1,2\}\,,
\label{A+} 
\end{equation}
i.e. on $\A_+$ the functions $R_i$ may be interpreted as two kinds of mean values of 
$\be_1$ and $\be_2$. 
Again beware that for 
$\be_2<0$ we may have $R_i<0$ even on $\A_+$. 
To explain the meaning of $R_0$ note that for $a>0$ the value of the replacement number $X$ at the  disease-free equilibrium (DFE)
of the RN-dynamical system \eqref{dot_X}-\eqref{dot_Y} is precisely given by $X_0^*=R_0$. 
Following results of \autocite{Driesche_Watmough2002} this leads to 

{\definition\label{Def_R0} 
$R_0$ is called the {\em reduced reproduction number}.
}

{\remark
As has been shown by \autocite{Driesche_Watmough2002,
Driesche_Watmough2008}, in models with just one infectious compartment the more general notion of $\R_0$ as the spectral radius of the next generation matrix (\autocite{Diekmann_et_al}, see also \autocite{DiekmannHeesterbeek}) reduces to the above definition. Denoting the values of $S_i$ at the DFE by
$S_i^*$ we have $R_0=\be_1S_1^*+\be_2S_2^*$, which is the usual formula, see e.g.  \autocite{KribsVel} or
\autocite{AvramAdenane2022}.
}
%
%if $R_0$ denotes the value of the replacement number at the DFE then under quite general assumptions (satisfied in particular by our present SSISS model, see Section \ref{Sec_EP}) the DFE is a stable node for $R_0<1$ and becomes a saddle point for $R_0>1$. 

{\remark
Mostly in the literature $R_0$ is called the {\em basic} reproduction number. But in case 
$\beta_2=0$ this terminology is already occupied by $r_0:=\beta_1/\ga$ as the expected number of secondary cases produced by a typical infectious individual in a totally susceptible population. So to avoid confusion I prefer to call $R_0$ the {\em reduced} reproduction number.
}

\bsn
Next put 
$\D_x:=\pi_\D(\phi(\A_x))$, $x=\mathrm{phys}$ or $x=\mathrm{bio}$, where $\pi_\D:\D\times\B\rto\D$ denotes the canonical projection. 
We look for suitable coordinates describing $\D_x$ and then derive additional bounds on $\bbe$ to describe
$\phi(\A_x)$. Consider the following functions on $\D$.
\begin{align}
A_\pm(\bx)	
&:=\frac{1}{2}\left(a+c\pm\sqrt{(a+c)^2-4(b+\ep)}\right),\label{A_pm_def} \\
 B_\pm(\bx)
&:=\frac{1}{2}\left(c\pm\sqrt{c^2-4\ep}\right)\,.
\label{B_pm_def} 
\end{align} 
Then by \eqref{normal_matrix} and the trace-det formula $A_\pm$ and $B_\pm$ provide the eigenvalues of $E+L$ and $L$, respectively.
The meaning of these  eigenvalues becomes clear by looking at \eqref{inv_param} 
\begin{align}
\be_1=A_+ 	& \LRA\al_1+\theta_1=0	&
\be_1= B_+	& \LRA\theta_1=0
\label{constraint_theta1} \\
\be_2=A_-	& \LRA\al_2+\theta_2=0	&
\be_2= B_-	& \LRA\theta_2=0
\label{constraint_theta2} \\
\be_i=R_0\,	& \LRA\al_i=0			&
\be_i=R_1	& \LRA\ga_j=0,\ j\neq i
\label{constraint_al+ga} 
\end{align}
More generally from \eqref{inv_param} we get
\begin{align}
\theta_i	&=\frac{(\be_i-B_-)(\be_i-B_+)}{\be_j-\be_i}\,,
\qquad j\neq i\,,\label{theta=be_i-Bpm}\\
\al_i+\theta_i	&=\frac{(\be_i-A_-)(\be_i-A_+)}{\be_j-\be_i}\,,
\qquad j\neq i\,.\label{alpha+theta}
\end{align}
Hence $A_\pm$  will serve to fix the constraints on 
$(\bx,\bbe)\in\phi(\Aph)$ 
 and $B_\pm$ ($B\equiv$ ``bio'') to fix constraints on 
 $(\bx,\bbe)\in\phi(\Abio)$. First we gather some trivial identities.
\begin{align}
c&=B_++B_-=A_++A_--a\,,	& \ep	&=B_+B_-=A_+A_--b\,,\label{AB_pm_variable}\\
a&=A_++A_--B_+-B_-\,,	& aR_0	&\equiv b =A_+A_--B_+B_-
\label{aR_0}
\end{align} 
From these one immediately computes
\begin{align}
a(A_\pm-R_0)	&=(A_\pm-B_+)(A_\pm-B_-) =
A_\pm^2-cA_\pm +\ep	\label{A-R_0}\\
a(R_0-B_\pm)	&=(B_\pm-A_+)(B_\pm-A_-) =
B_\pm^2 -(a+c)B_\pm +(b+\ep)	\label{B-R_0}
\end{align}
Now let's introduce the notation
\begin{align}
\D_A	&:=\D\cap\{A_\pm\in\RR\}\label{DA}\\
\D_B	&:=\D\cap\{B_\pm\in\RR\land B_-<B+\}\label{DB}\\
\DAB	&:=\D_A\cap\D_B\cap\{B_-\leq A_-\leq B_+\leq A_+\}.
\label{DAB}
\end{align}

{\lemma\label{Lem_DAB}
The following identities hold
\begin{equation}
\DAB =\{\bx\in\D_A\mid A_-\leq R_0\leq A_+\,\land\, c^2\neq 4\ep\}
=\{\bx\in\D_B\mid B_-\leq R_0\leq B_+\}	\label{B-<R_0<B+}
\end{equation}
Hence in $\DAB$ we always have the additional bound
\begin{equation}
B_-	\leq A_- \leq R_0 \leq B_+ \leq A_+\,.
\label{A_-<R_0}
\end{equation}
}
\begin{proof}
By Eqs. \eqref{A-R_0} and \eqref{B-R_0} on $\DAB$ we always have
$A_-\leq R_0\leq A_+$ and $B_-\leq R_0\leq B_+$. Conversely, if
$\A_\pm\in\RR$, $c^2\neq\ep$ and $A_-\leq R_0\leq A_+$ then \eqref{A-R_0} implies 
$A_-^2-cA_-+\ep\leq 0\leq A_+^2-cA_++\ep$ and therefore 
$c^2> 4\ep$. Hence 
$B_-<B_+\in\RR$ and again by \eqref{A-R_0} 
$B_-\leq A_-\leq B_+\leq A_+$. 
%Now use \eqref{B-R_0} to conclude $B_-\leq R_0\leq B_+$.
The second identity follows analogously.
\end{proof}
%\begin{equation}
%\D_A=
%\{\bx=(a,R_0,R_1,A_+,A_-)\in\RR_+\times\RR^4\mid
%A_-\leq A_+\}.
%\label{D_P} 
%\end{equation}
%
%\begin{equation}
%c= B_++ B_-\,,\qquad\ep= B_+ B_-\,.\label{AB_pm_variable} 
%\end{equation}
%to reparametrize $\D_B$ as
%\begin{equation}
%\D_B=
%\{\bx=(a,R_0,R_1,B_+,B_-)\in\RR_+\times\RR^4\mid B_-< B_+\}.
%\label{D_B} 
%\end{equation}
%Putting $\DelB:=\B_+-B_-\equiv\sqrt{c^2-4\ep}$ we also denote 
%\begin{equation}
%\DBo:=\D_B\cap\{\DelB>0\}.
%\label{DB_Del}
%\end{equation}

{\lemma\label{Lem_P+B} 
Denoting  $i_B:\D_B\ni\bx\mapsto (\bx,B_+,B_-)\in\D_B\times\B$ 
the following identities hold
\begin{align}
\phi(\Asig)	&=\{(\bx,\bbe)\in\D_B\times\B\mid
B_-\leq\be_2<\be_1\leq B_+\}\,,
\label{A_sig} \\
\phi(\A_{\bth+\bal\geq 0})	&=\{(\bx,\bbe)\in\D_A\times\B\mid
\be_2\leq A_-\leq\be_1\leq A_+\}\,.
\label{A_theta+alpha}\\
\phi(A_{\bth=0})	&=i_B(\D_B)\label{i_B(D_B)}
\\
\phi(A_{\bth=0}\cap\A_{\bal\geq 0}) &=i_B(\DBA)	\label{i_B(D_BA)}	
\end{align}
}
\begin{proof}
We have $B_\pm\in\RR$ iff there exists $\be\in\RR$ such that 
$\be^2-c\be+\ep\leq 0$. Hence, by \eqref{inv_param}, if 
$\theta_1\geq 0$ and $\theta_2\leq 0$ then $B_\pm\in\RR$ and 
$B_-\leq\be_2<\be_1\leq B_+$, proving the ``$\subset$''-part in 
\eqref{A_sig}. The opposite direction follows from \eqref{theta=be_i-Bpm}. Similarly, $A_\pm\in\RR$ iff there exists $\be\in\RR$ such that $\be^2-(a+c)\be+b+\ep\leq 0$. Hence, by \eqref{inv_param}, if $\al_1+\theta_1\geq 0$ then $A_\pm\in\RR$ and $A_-\leq\be_1\leq A_+$. If in addition $\al_2+\theta_2\geq 0$ then also $\be_2\leq A_-$, proving the ``$\subset$''-part in \eqref{A_theta+alpha}. The opposite direction follows from 
\eqref{alpha+theta}.
\Eqref{i_B(D_B)} follows since in $\A_{\bth=0}$ we have 
$\be_1=B_+$ and $\be_2=B_-$.
If in addition $\al_i\geq 0$ then \eqref{alpha+theta} implies \Eqref{i_B(D_BA)}. 

%Finally, 
%$\al_i\geq 0\LRA R_0\in[\be_2,\be_1]$ by \eqref{inv_param}, thus proving \Eqref{DAB}.
%%
%To prove part ii), 
%\eqref{a(P,B)} follows from Eqs.  \eqref{AB_pm_variable} and 
%the definitions \eqref{A_pm_def} - \eqref{B_pm_def} straight forwardly yield
%\eqref{A-R_0} - \eqref{B-R_0} and consequently also 
%\eqref{A_-=R_0} - \eqref{B_+=R_0}.
%Finally, if $\bx\in\DBA$ then $B_-\leq A_-\leq B_+$ together with \Eqref{A-R_0} proves $A_--R_0\leq 0$ and hence the remaining inequality in \Eqref{A_-<R_0}. 
%
%assume $\bx\in\D_B$ and  $B_-\leq R_0\leq B_+$.
%Plugging \Eqref{AB_pm_variable} into \eqref{A_pm_def} and using  $b=aR_0$ gives
%$A_\pm=(a+ B_-+ B_+\pm\sqrt{D})/2$
%where 
%$$D:=(a+ B_-+ B_+)^2-4(aR_0+ B_- B_+)=a^2+(B_+-B_-)^2+2a(B_++B_--2R_0).
%$$
%Now use $(B_--R_0)^2\leq(B_+-B_-)^2$ and
%$
%B_--B_+\leq B_--R_0\leq B_++B_--2R_0\leq B_+-B_-
%$
%to conclude
%\begin{equation}
%\max\{(a+ B_-- B_+)^2,(a+ B_--R_0)^2\}\leq D\leq (a+ B_+- B_-)^2.
%\label{bounds}
%\end{equation}
%Hence $D\geq 0$ implying $\bx\in D_P$  and the bounds \eqref{bounds} straightforwardly lead to
%\begin{align*}
%B_-	&\leq A_- \leq\min\{B_-+a,(B_++R_0)/2\}\leq B_+\\
%B_++a	&\geq A_+ \geq \max\{B_+,B_++B_-+a-R_0\}
%\end{align*}
%implying  \eqref{A_-<R_0} with $R_0$ being replaced by 
%$(B_++R_0)/2$.
%Now use $\A_--R_0\leq 0$ by \eqref{a(P,B,R_0)} and 
%$R_0\leq(B_++R_0)/2$  to get \eqref{A_-<R_0}.
%
%Finally, \Eqref{A_-=R_0} follows from by \eqref{a(P,B,R_0)} and also by  \eqref{a(P,B,R_0)} $B_+=R_0$ iff $A_+=B_+$ or 
%$a=A_+-B_-$. The later being equivalent to $B_+=A_-$ by \eqref{a(P,B)} proves \eqref{B_+=R_0}.
\end{proof}

\bsn
We are now in the position to summarize the constraints describing 
$\phi(\Aph)$ and $\phi(\Abio)$.

{\proposition\label{Prop_Aph+Abio}
For $\A_x=\C_x\times\B$ as defined in \eqref{Cph} - \eqref{Cbio} we have
\begin{align}
\phi(\Aph)	&\equiv\phi(\A_+)\cap\phi(\A_{\bth+\bal\geq 0})
\notag\\
&=(\D_A\times\B)\cap
\{\be_2\leq \{A_-,R_0,R_1\}\leq\be_1\leq A_+\}\,,
\label{A_ph}\\
\Dph	&=\D_A\cap\{R_{0,1}\leq A_+\}\,,
\label{D_ph}\\
\phi(\Abio)	&\equiv\phi(\Asig)\cap\phi(\Aph)\notag\\
		&=(\DBA\times\B)\cap
\{B_-\leq\be_2\leq A_-\leq R_0\leq\be_1\leq B_+\}\cap
\{R_1\in[\be_2,\be_1]\}\,,
\label{A_bio}\\
\Dbio	&=\DBA\cap\{R_1\in[B_-,B_+]\}\subset\Dph\,.
\label{D_bio}
\end{align}
}
\begin{proof}
This is a summary of 
\Eqref{A+} and Lemmas \ref{Lem_DAB} - \ref{Lem_P+B}.
\end{proof}

\noindent
Proposition \ref{Prop_Aph+Abio} motivates the following notation and definition

{\definition\label{Def_compatible}
For $\bx\in\Dbio$ put
\begin{equation}
\be_2^{\max}(\bx):=\min\{A_-,R_1\}, \qquad
\be_1^{\min}(\bx):=\max\{R_0,R_1\}.
\label{beta_min_max}
\end{equation}
Then $\bbe\in\B$ is called 
{\em bio-compatible} with $\bx$ if $B_-\leq\be_2\leq\be_2^{\max}$ and $\be_1^{\min}\leq\be_1\leq B_+$, equivalently if $\phi\inv(\bx,\bbe)\in\Abio$. Similarly, $\bbe$ is called 
{\em compatible} if $\be_2\leq\be_2^{\max}$ and 
$\be_1^{\min}\leq\be_1\leq A_+$, equivalently if $\phi\inv(\bx,\bbe)\in\Aph$. A physical triangle $\Tph(\bbe)$ is called (bio)-compatible, if $\bbe$ is (bio)-compatible.
}

\bsn
Hence, (bio-)compatible physical triangles are always forward invariant under the RN-dynamics \eqref{dot_X}-\eqref{dot_Y} and the smallest one is just $\Tph(\be_1^{\min},\be_2^{\max})$.
The following Corollary also proves part ii) of Theorem \ref{Thm_periodic}.

\begin{corollary}\label{Cor_periodic}
Let $\bx\in\Dbio$ and let $\bbe\in\B$ be compatible with $\bx$.
Then there exist no periodic solutions, homoclinic loops or oriented phase polygons of the RN-dynamical system \eqref{dot_X}-\eqref{dot_Y} in $\Tph(\bbe)$.
\end{corollary}
\begin{proof}
Let $\Z\subset\Tph(\bbe)$ be a solution cycle (image of a periodic solution, a homoclinic loop or an oriented phase polygon). As argued in the proof of  Theorem \ref{Thm_periodic}, we must have 
$\Z\neq\partial\Tph(\bbe')$ for all $\bbe'\in\B$. Hence, by forward invariance, 
$\Z$ must lie inside the smallest compatible triangle, 
$\Z\subset\Tph(\be_1^{\min},\be_2^{\max})\subset\Tph(B_+,B_-)$. But, by Proposition \ref{Prop_Aph+Abio} and \Eqref{i_B(D_BA)},
$\phi\inv(\bx,B_+,B_-)\in\Abio\cap\A_{\bth=0}$ and we get a contradiction with Theorem \ref{Thm_periodic}i).
\end{proof}

\noindent
Finally, to prove Theorem \ref{Thm_gaugefix}v), note that Lemma  
\ref{Lem_P+B} and Proposition \ref{Prop_Aph+Abio} in particular imply (use that $G_S$ acts transitively on $\B$)
\begin{align}
A_{\bth=0}\li G_S=\Asig\li G_S	&=\phi\inv(\D_B\times\B)
\label{orbit_Athe=0}\\
\A_{\bth+\bal\geq 0}\li G_S	&=\phi\inv(\D_A\times\B)
\\
(A_{\bth=0}\cap\A_{\bal\geq 0})\li G_S &=\phi\inv(\DBA\times\B)
\\
\Aph\li G_S &= \phi\inv(\Dph\times\B)
\\
\Abio\li G_S &= \phi\inv(\Dbio\times\B)
\\
\A_{\bth=0}\cap\Abio\li G_S &\subset\Abio
\label{orbit_Abio}
\end{align}
where the last equation follows from $\A_{\bth=0}=i_\B(\D_B)$ and
$i_\B(\D_B)\cap(\Dbio\times\B) =i_B(\Dbio)\subset\phi(\Abio)$.
Theorem \ref{Thm_gaugefix}v) now follows from Eqs.
\eqref{orbit_Athe=0}, \eqref{orbit_Abio} and Lemma \ref{Lem_A_B}
below.
\begin{lemma} \label{Lem_A_B} 
Put $\A_B:=\phi\inv(\D_B\times\B)=A_{\bth=0}\li G_S$, then
$$\A_B\supset\A\cap\{\theta_1\geq\theta_2\,\lor\,\theta_1\theta_2>0\}\supset\Asig\supset\Abio.$$
\end{lemma}
\begin{proof}
The second and third inclusions are obvious from the definitions 
\eqref{Csig} and \eqref{Cbio} and the first inclusion follows from
$\D_B=\D\cap\{c^2>4\ep\}$ and
\begin{align*}
c^2-4\ep&=(\be_1-\be_2)^2+(\theta_1+\theta_2)^2+
2(\be_1-\be_2)(\theta_1-\theta_2)\\
&=(\be_1-\be_2+\theta_1-\theta_2)^2+4\theta_1\theta_2.
\end{align*}
\end{proof}

\begin{table}[htpb!]
\caption{Sector classification in $\Abio$ generalizing Table \ref{Tab_1}.}
\label{Tab_1'} 
\renewcommand{\arraystretch}{1.4}
\begin{tabular}{|c|c|c|l|}
\hline
	Sector & $c=B_-+B_+$ & 
	$\ep=B_-B_+$ & 
	Interval $[B_-,B_+]$ 
\\
\hline
\RN{1}&$+$& $+$& $0<B_-<B_+$  
\\\hline
\RN{2} (SIRS)&$+$& $0$& $0=B_-<B_+$ 
\\\hline
\RN{3}&$+$& $-$& $0<-B_-<B_+$ 
\\\hline
\RN{4}&$0$& $-$& $0<-B_-=B_+$ 
\\\hline
\RN{5}&$-$& $-$& $B_-<-B_+<0$ 
\\\hline
\RN{6}&$-$& $0$& $B_-<B_+=0$ 
\\\hline
\RN{7}&$-$& $+$& $B_-<B_+<0$ 
\\\hline
\end{tabular}
\renewcommand{\arraystretch}{1}
\end{table}
\bsn
Let me close by mentioning that the parametrizations \eqref{AB_pm_variable} can now be used to generalize the Sector classification of Table \ref{Tab_1} from the special case 
$\A_{\bth=0}$ to all of $\Abio$ (more generally to
$\A_B:=\phi\inv(\D_B\times\B)\supset\Abio$) as shown in Table \ref{Tab_1'}.

\subsection{Examples revisited\label{Sec_examples2}}
For completeness let us revisit the examples in Section \ref{Sec_examples} within the present setting. Eqs. \eqref{Heth}-\eqref{AABH} translate into\footnote{
${\mathrm{Heth}}=$ \autocite{Hethcote1974, Hethcote1976, Hethcote1989};
${\mathrm{SIRI}}=$ \autocite{DerrickDriessche};
${\mathrm{BuDr}}=$ \autocite{BusDries90};
${\mathrm{SIRS}}=$ 10-parameter mixed SIRS/SIS model with constant population size and $\theta_2=\be_2=0$;
${\mathrm{HaCa}}=$ core system in \autocite{Had_Cast};
${\mathrm{KZVH}}=$ \autocite{KribsVel};
${\mathrm{LM}}=$ \autocite{LiMa2002};
${\mathrm{AABH}}=$ \autocite{AvramAdenane2022}.
${\mathrm{SIRI}}$ and ${\mathrm{AABH}}$ come in two versions, the subscript $1$ refers to 
$\be_S>\be_R$ and $2$ to $\be_S<\be_R$.
}
\begin{align}
\D_{\mathrm{Heth}}	&=\Dbio\cap
\{R_0=B_+\,\land\,a<1\,\land\,d=B_-=0\}
%\cap\{R_1<R_0=B_+\,\land\,B_-=0\,\land\,a=R_1/B_+\}
\label{Heth2}\\
\D_{\mathrm{SIRI}_{1,2}} &=\Dbio
\cap\{R_0=B_\pm\,\land\,a<1\,\land\,d=B_\mp(B_\pm+1-a)\} 
\label{SIRI2}\\
\D_{\mathrm{BuDr}} &=\Dbio
\cap\{R_1<R_0=B_+\,\land\,B_-<0\}
\footnotemark\addtocounter{footnote}{-1}
\label{BuDr2}\\
\Dsirs	&=\Dbio\cap\{B_-=0\}
\label{SIRS2} \\
\D_{\mathrm{LM}} &=\Dbio\cap\{B_-<\min\{0, R_1\}\}
\label{JLZM2}\\
\D_{\mathrm{KZVH}} &=\Dbio\cap\{B_->0\}
=\D_{\mathrm{HaCa}}
\label{KZVH2}\\
\D_{\mathrm{AABH_1}} &=\Dbio\cap\{B_-\leq R_1< B_+\}
\footnotemark\addtocounter{footnote}{-1}
\label{AABH_1}\\
\D_{\mathrm{AABH_2}} &=\Dbio\cap\{B_-< R_1\leq B_+\}
\footnotemark
\label{AABH_2}
\end{align} 

\noindent
Note that all models except SI(R)S already satisfy 
$\theta_i=0$ whence $\tilde{\be}_1=B_+$, $\tilde{\be}_2=B_-$ by Eqs. \eqref{constraint_theta1}-\eqref{constraint_al+ga}. In the SI(R)S model we have instead 
$0=\tilde{\be_2}=B_-<\tilde{\be_1}\leq B_+$.
Corollary \ref{Cor_Avr=JiLi+KrZa} may now be reformulated as follows
\begin{corollary}\label{Cor_Avr=JiLi+KrZa_2}
Referring to the sub-cases $\mu_1=\mu_2$ in \autocite{BusDries90,
AvramAdenane2022} and putting
$\D_{\mathrm{AABH}}:=\D_{\mathrm{AABH}_1}\cup\D_{\mathrm{AABH}_2}$
we have
\begin{align}
\D_{\mathrm{Heth}}	&=\D_{\mathrm{SIRI}_1}\cap\{B_-=0\}
\\
&=\Dsirs\cap\{a<1\,\land\,R_0=c\,\land\,d=0\}
\label{Heth-SIRI}\\
\D_{\mathrm{LM}} &\supset\D_{\mathrm{BuDr}}\cap\{B_-\neq R_1\}
\label{BuDr-LM}\\
\D_{\mathrm{LM}} &=\D_{\mathrm{AABH_2}} \cap\{B_-<0\}
\label{LM-AABH}\\
\D_{\mathrm{KZVH}} &=\D_{\mathrm{AABH}} \cap\{B_->0\}
\label{KZVH-AABH}
\end{align}
\end{corollary}
\footnotetext{Referring to the sub-case $\mu_1=\mu_2$ in these models, see Footnotes \ref{FN_AABH} and \ref{FN_BuDr}.}

\noindent
Finally, we are now  in the position to generalize the scaling symmetry for SI(R)S models of \autocite{Nill1} to the present setting. First note that having started from the 10-parameter extended SI(R)S model we now have arrived at $\dim\Dsirs =4$. Also, 
$\dim\D_{\mathrm{Heth}}=2$ with independent parameters $a\in(0,1)$ and $c=R_0=B_+>0$. In particular, if 
$\bx\in\D_{\mathrm{Heth}}$ then putting $(u,v):=(X,cI)$ the RN-dynamical system \eqref{dot_X}-\eqref{dot_Y}  reduces to the classic endemic model in \Eqref{normalization}. In a second normalization step the number of parameters in the SI(R)S case may now be reduced again by two.
In this way, for $c>d$,\footnote{Note that in $\Dsirs$ we have $c=B_+\geq R_1-aR_0=d$ where equality implies $R_0=0$ and 
$R_1=B_+$.}
the normalized SI(R)S model also looks like the classic endemic model 
\begin{equation}
\dot{u}=-uv-c_1u+c_2\,,\qquad\dot{v}=uv-v\,,
\label{class_end}
\end{equation}
the difference being that coming from $\D_{\mathrm{Heth}}$ we have $c_1=a\in(0,1)$ and $c_2=aR_0\geq 0$, whereas coming from $\Dsirs$ gives
$(c_1,c_2)\in\RR_+\times\RR$\footnote{In case $a=0$ we would get $c_1=c_2=0$.}. 
%In this way one gets
%
%\begin{theorem}\label{Thm_SIRS}
%Let $\bx\in\Dsirs\cap\{0<a<1-d/c\}$. Then there exists 
%$\bx'\in\D_{\mathrm{Heth}}$ such that the RN-dynamical systems 
%\eqref{dot_X}-\eqref{dot_Y} at parameter values $\bx$ and $\bx'$ are isomorphic.
%\end{theorem}
%\begin{remark}
%The restriction to $a<1-d/c$ in Theorem \ref{Thm_SIRS} is needed to obtain $c_2>0$. 
However, since endemic bifurcation in the model \eqref{class_end} occurs at $R_0=c_2/c_1=1$, extending this model to the SI(R)S case by including also values $c_2<0$ and $c_1\geq 1$ doesn't change its characteristic behavior. In particular, various proofs in the literature on variants of constant population SI(R)S models with standard incidence become obsolete, it's all contained in Hethcote's work.

%\end{remark}
%\noindent
%Theorem \ref{Thm_SIRS} and 
\bsn
\Eqref{class_end} is proven in 
Appendix \ref{Sec_SIRS}. In principle, the proof relies on the same structure as in Theorem \ref{Thm_gaugefix}, with the symmetry group $G_S$ acting on $\A$ replaced by a dilatation group 
$\Gdil=\RR_+^2$ acting on $\D$. Since these dilatations may blow up physical triangles to arbitrary size, we also get the following

\begin{lemma}\label{Cor_bounded}
For $\bx\in\Dsirs$ the forward flow of the RN-dynamical system 
\eqref{dot_X}-\eqref{dot_Y} stays bounded for all initial conditions $(X_0,I_0)\in\RR\times\RRN$.
\end{lemma}
\noindent
This result may be used to prove, that SI(R)S models as above are always Hamiltonian \autocite{Nill5}. Lemma \ref{Cor_bounded} is also proven in Appendix \ref{Sec_SIRS}.

%
%\bsn
%Let close with generalizing the result  by showing that in the SI(R)S case there is a second normalization step, reducing the number of parameters again by two. First note that for $\bx\in\D_{\mathrm{Heth}}$ we may put $(u,v)=(X,cI)$ to obtain the fully normalized version of the classic endemic model, see \Eqref{normalization}.
%
%\begin{equation}
%\dot{u}=-uv-a(u-R_0),\qquad\dot{v}=uv-v,\qquad,0<R_0,\,0<a<1.
%\end{equation}
%Putting $(x,y)=(u-1,v)$ we get equivalently
%\begin{equation}
%\dot{x}=-(x+1)y+\ka_0-\ka_1y, \qquad\dot{y}=xy,\qquad,0<R_0,\,0<a<1.
%\end{equation}

\section{Summary and outlook \label{Sec_Summary}}
In summary we have seen, that in canonical coordinates 
the 14-parameter SSISS model, constraint by $\nu_1=\nu_2$, effectively depends on at most five parameters 
$\bx=(a,b,c,d,\ep)$. Depending on natural model restrictions 
like ``$\mathrm{phys}$'' or ``$\mathrm{bio}$'' these parameters obey various relations which can be encoded by further reparametrizations like $\bx=(a,R_0,R_1,B_+,B_-)$, see Eqs. \eqref{R_0}, \eqref{R_1}, \eqref{AB_pm_variable} and Proposition \ref{Prop_Aph+Abio}. The incidence rates
$\be_i$ have disappeared from the equations of motion. Their role  is reduced to fixing physical triangles $\Tph(\bbe)$ in 
$(X,I)$-space, see \Eqref{Tph2}. If
$\bx\in\Dbio$, then for all compatible values $\bbe=(\be_1,\be_2)$   the triangles
$\Tph(\bbe)$ stay forward invariant under the RN-dynamics \eqref{dot_X}-\eqref{dot_Y}. 
Independence of $\bbe$ also means that SSISS models at parameter values $\phi\inv(\bx,\bbe)$ for fixed 
$\bx\in\D$ and varying 
$\bbe\in\B$ are all 
isomorphic to each other\footnote{By Remark \ref{Rem_phys-Tr}, physical triangles are not mapped onto each other under these isomorphisms.} (Proposition \ref{Prop_SS'}). The isomorphisms are provided by a parameter symmetry group $G_S\subset GL_+(\RR^2)$ acting simultaneously on phase space $\P$ and parameter space 
$\A$ (Theorem \ref{Thm_gaugefix}i-iv).
If $\bx\in\D_B$ then a representative in $\A$ of the equivalence class $\bx$ may always be chosen by putting $\be_1=B_+$ and 
$\be_2=B_-$ and hence $\theta_i=0$ (Theorem \ref{Thm_gaugefix}v). In combination with methods from \autocite{BusDries90} this also leads to a proof of absence of periodic solutions for all $\bfa\in\Abio$ (Theorem \ref{Thm_periodic}).

%By convenient abuse of physics terminology I will call such model classes {\em gauge equivalent}.
\bsn
In part III of this work it will be shown, that the model also admits an additional scaling symmetry leading to a second normalization step, similar as described for the SI(R)S model in
Appendix \ref{Sec_SIRS}, see also
\autocite{Nill1}. In this way the number of essential parameters will further reduce from five to three (respectively two in Sectors \RN{2} and \RN{6}).

\bsn
Part II of this work will reanalyze equilibrium points and their stability properties in all Sectors of $\Abio$, thereby recovering and extending the results of 
\autocite{Had_Cast, KribsVel, LiMa2002, AvramAdenane2022}, which had been obtained for $\theta_i=0$ and some more parameter restrictions, see Table \ref{Tab_examples} and  Corollary  \ref{Cor_Avr=JiLi+KrZa}/\ref{Cor_Avr=JiLi+KrZa_2}.
This approach will differ from previous papers by relying on the normalization formalism and sector classification of the present work. In this way the search for endemic equilibria 
$(X^*,I^*)$ simplifies considerably, since always $X^*=1$. So one is left with analyzing roots of the quadratic equation 
$h(I^*):=\dot{X}(X^*=1,I^*)=0$. 
This will also uncover an exceptional scenario in Sectors III-V, which apparently has been overlooked in the literature so far.

%
%
%For $R_0<1$ the so-called backward-bifurcation scenario  (i.e. coexistence of the disease free equilibrium with two endemic equilibria) will show up in a certain sub-region in parameter space of Sector \RN{1}. In all other Sectors the disease free equilibrium will turn out to be unique and globally stable provided $R_0\leq 1$. Note that 
%$R_0<1$ (in fact $R_0\leq 0$) always holds in Sectors \RN{6}-\RN{7}. 
%
%If $R_0>1$  then the disease free EP becomes a saddle and there always exists a locally asymptotically stable endemic equilibrium  
%$\bP^*\in\Tph(\be_1^{\min},\be_2^{\max})$. Outside an exceptional set in parameter space in Sectors \RN{3}-\RN{5}, $\bP^*$ is unique and globally stable away from $\{I=0\}$ in all compatible physical triangles. In the exceptional parameter range there exists a second endemic equilibrium in the boundary 
%$\partial\Tph(\be_1^{\min},\be_2^{\max})$, which turns out to be a saddle with attractive line given by 
%$\{S_2=0\}$. In this case $\bP^*$ stays globally stable away from 
%$\{I=0\}\cup\{S_2=0\}$. Using the mapping in Table \ref{Tab_examples} and the identifications \eqref{LM}-\eqref{AABH}, the exceptional case should also show up in the setting of \autocite{LiMa2002} and \autocite{AvramAdenane2022}, which apparently had been overlooked by the authors.

\appendix

\section{Normalizing linear vital dynamics \label{Sec_vital-dynamics}}
This Appendix gives a normalization prescription for the dynamics of fractional variables in an $n$-compartment model with linear vital dynamics. Let the vectorfield 
$\bV:\RR^n\rto\RR^n$ be homogeneous of degree one and assume there exists $\bnu=(\nu_1,\cdots,\nu_n)$ such that 
$\bra\one|\bV(\bY)\ket\equiv\sum_i V_i(\bY)=\bra\bnu|\bY\ket$ for all $\bY\in\RR^n$, where $\one:=(1,\cdots,1)$. Call 
$N(\bY):=\bra\one|\bY\ket$ the total population and 
$\by:=N\inv\bY$ the fractional compartment variables, then the dynamical system 
$\dot{\bY}=\bV(\bY)$ implies
$$
\dot{\by}=\bV(\by)- \bra\bnu\mid\by\ket \by=:\bF(\by).
$$
Denote $\S:=\{\by\in\RR^n\mid\bra\one|\by\ket=1\}$, then clearly 
$\bra\one|\bF\ket|_\S=0$. The aim is to substitute $\bF$ by 
$\tilde{\bF}$ such that 
$\bF|_\S=\tilde{\bF}|_\S$ and $\bra\one|\tilde{\bF}\ket=0$ holds as an identity on all of $\RR^n$. The following Lemma holds by straight forward calculation.

\begin{lemma}\label{Lem_tildeF}
Put $\Lambda_{ijk}:=(\delta_{ij}-\delta_{ik})(\nu_k-\nu_j)$ and
$\Lambda_i(\by):=\sum_{j,k}\Lambda_{ijk}y_jy_k$.
\begin{itemize}
\item[i)]
For all $\by\in\RR^n$ and $i=1,\cdots,n$ we have
\begin{equation}
\frac{1}{2}\Lambda_i(\by)=\sum_k(\nu_k-\nu_i)y_iy_k
\equiv y_i\bra\bnu|\by\ket-\nu_iy_i\bra\one|\by\ket.
\label{Lambda}
\end{equation}
\item[ii)]
Put 
\begin{equation}
\tilde{\bF}:=\bV-\diag(\bnu)-\frac{1}{2}\bLa.
\label{tilde(F)}
\end{equation}
Then $\bF|_\S=\tilde{\bF}|_\S$ and $\bra\one|\tilde{\bF}\ket=0$ as an identity on $\RR^n$.
\end{itemize}
\end{lemma}

\bsn
By this method we also get conditions guaranteeing that 
%Let me conclude with demonstrating that in this way in models where $\bV$ contains certain linear and quadratic terms 
constant per capita birth and death rates become redundant as in \Eqref{notation2}. 
\begin{lemma}
Let $\bV(\bY)$ be of the form
$$
V_i(\bY)=\sum_j M_{ij}Y_j+\frac{1}{2}\sum_{j,k}\Gamma_{ijk}Y_jY_k/N + 
\sum_j L_{ij}Y_j
$$
where without loss 
$\Gamma_{ijk}=\Gamma_{ikj}$ and where $\sum_iM_{ij}=\sum_i\Gamma_{ijk}=0$. Hence, all vital dynamics parameters are encoded in $(L_{ij})$ and $\nu_j:=\sum_iL_{ij}$ satisfies 
$\bra\one|\bV=\bra\bnu|$. If in this case
$L_{ij}\neq \nu_i\delta_{ij}\Rightarrow M_{ij}\neq 0$ and 
$\nu_j\neq\nu_k\Rightarrow 
(\Gamma_{jjk}\neq 0\,\land\, \Gamma_{kkj}\neq 0)$, then for the dynamics of fractional variables all parameters $L_{ij}$ are redundant.
\end{lemma}
\begin{proof}
Applying \eqref{tilde(F)} we have 
$\tilde{F}_i(\by)=\sum_j \tilde{M}_{ij}y_j+
\frac{1}{2}\tilde{\Gamma}_{ijk}y_jy_k$,
where $\tilde{M}_{ij}=M_{ij}+L_{ij}-\nu_i\delta_{ij}$ and 
$\tilde{\Gamma}_{ijk}=\Gamma_{ijk}-\Lambda_{ijk}$. The claim follows since $\Lambda_{ijk}=\Lambda_{ikj}$, $\Lambda_{jjk}=-\Lambda_{kkj}$ and $\Lambda_{ijk}=0$ if $\nu_j=\nu_k$ or if 
$j\neq i\neq k$, which also yields $\sum_i\Lambda_{ijk}=0$ .
\end{proof}

\section{Scaling the SI(R)S model \label{Sec_SIRS}}

In this appendix we extend the dilatation symmetry as proposed for a 6-parameter SI(R)S model in \autocite{Nill1} to the 10-parameter extended SI(R)S model as classified in this paper. Denote Sector \RN{2} in $\D_B$ by
$\D_{\RN{2}}:=\D_B\cap\{B_-=0\}$ and  $\Dsirs:=\D_{\RN{2}}\cap\Dbio$. Recall that in $\D_{\RN{2}}$ we have $c=B_+>0$ and in $\Dsirs$ we have $0\leq R_i\leq B_+$ and hence $d-c=R_1-aR_0-B_+\leq 0$, where equality implies $R_0=0$ and $R_1=B_+$. Hence the following Lemma in particular includes Lemma \ref{Cor_bounded}.

\begin{lemma}\label{Lem_SIRS}
Consider the RN-dynamical system \eqref{dot_X} - \eqref{dot_Y} on phase space $\P\equiv\RR\times\RRN$ for parameter values 
$\bx=(a,b,c,d,\ep=0)\in\D_{\RN{2}}\cap\{d\leq c\,\land\,d=c\Rightarrow R_0< 1\}\supset\Dsirs$.
Let $\T\subset\P$ be a rectangular triangle with corners 
$\bT_\triangleleft=(X_\triangleleft,0)$, 
$\bT_\triangleright=(X_\triangleright,0)$ and
$\bT_\vartriangle=(X_\triangleleft,I_\vartriangle)$, where
$X_\triangleleft<X_\triangleright$. Call $\T$
compatible with $\bx$ if 
\begin{align*}
I_\vartriangle &=(X_\triangleright-X_\triangleleft)/c
\\
X_\triangleleft &\leq\min\{R_0,d/c\}
\\
R_0-X_\triangleleft &\leq I_\vartriangle\min\{c,(c-d)/a\}
\end{align*}
\begin{itemize}
\item[i)]
Then every $\bx$-compatible triangle $\T$ is forward invariant.
\item[ii)]
The forward flow for arbitrary initial conditions $(X_0,I_0)\in\P$ stays bounded.
\end{itemize}
\end{lemma}

\begin{proof}
To prove part i), the upper bounds on $X_\triangleleft$ imply 
$\dot{X}>0$ on the line 
$\{X=X_\triangleleft\}$. We are left to show 
$\dot{X}+c\dot{I}\leq 0$ on the hypotenuse $X(I)=X_\triangleleft+c(I_\vartriangle - I)$, 
$0\leq I\leq I_\vartriangle$.
\begin{align*}
\dot{X}+c\dot{I}&=a(R_0-X(I))+(d-c)I
\\
&=a(R_0-X_\triangleleft-c(I_\vartriangle - I))+(d-c)I
\\
&\leq 
I_\vartriangle \min\{ac,c-d\}-ac(I_\vartriangle - I)
+(d-c)I
\\
&\leq 0
\end{align*}
Part ii) follows since for $d<c$ we may always choose 
$X_\triangleleft<X_0$ and 
$X_\triangleright$ large enough, such $\T$ is $\bx$-compatible and $(X_0,I_0)\in\T$. For $d=c$ and $R_0<1$ $\bx$-compatibility requires $X_\triangleleft=R_0$. If in this case $X_0<R_0$ glue the rectangle $\R=[X_0,R_0]\times[0,I_\vartriangle]$ to the left of 
$\T$. Then $(X_0,I_0)\in \R\cup\T$ for $X_\triangleright$ large enough and $\R\cup\T$ is forward invariant, since 
$\dot{I}< 0$ and $\dot{X}>0$ for $(X,I)\in\R$.
\end{proof}
\noindent
Given 
$\bx\in\D_{\RN{2}}\cap\{d\leq c\,\land\,d=c\Rightarrow R_0< 1\}$ as above and $\T$ compatible with $\bx$ we now show that the RN-dynamical system \eqref{dot_X} - \eqref{dot_Y} may always be rescaled to an isomorphic system with parameters 
$\bx'\in\Dsirs$ such that  $\T$  maps to the physical triangle 
$\Tph(B'_+,0)$ of the SI(R)S system. 
Following \autocite{Nill1} the 
{\em dilatation symmetry group}
$\Gdil\equiv\GX\times\GI\equiv\RR_+^2$ is defined by rescaling 
$(X,I)$ variables according to
$$
X_{(\xi,\la)}(t)-1:=\xi (X(\xi t)-1),\qquad
I_{(\xi,\la)}(t):=\la I(\xi t), \qquad
(\xi,\la)\in\RR_+^2
$$ 
The following Lemma is easily verified by straightforward calculation.
{\lemma
Let the group action 
$\re:\Gdil\times\D\ni(\xi,\la,\bx)\mapsto(\xi,\la)\re\bx\in\D$ be given by
\begin{equation}
(\xi,\la)\re(a,R_0-1,c,d-c,\ep):=
(\xi a,\xi(R_0-1),\xi c/\la,\xi^2(d-c)/\la,\xi^2\ep/\la^2)
\label{dil}
\end{equation}
and for $\bx\in\D$ let $\bff_\bx(X,I)$ denote the vector field of the system \eqref{dot_X} - \eqref{dot_Y}. Then 
$$
(\dot{X},\dot{I})=\bff_\bx(X,I)\Longleftrightarrow
(\dot{X}_{(\xi,\la)},\dot{I}_{(\xi,\la)})=
\bff_{\bx'}(X_{(\xi,\la)},I_{(\xi,\la)}),\qquad
\bx'=(\xi,\la)\re\bx.
$$
\qed}

\noindent
Note that this action leaves all Sectors in $\D_B$ invariant, but in general not $\Dbio\subset\D_B$. We now determine $\Gdil\re\Dsirs$, thereby also providing an alternative proof of
Lemma \ref{Lem_SIRS}i).

\begin{proposition}\label{Prop_scaling}\ 
\begin{itemize}
\item[i)]
Let $\T$ be compatible with
$\bx\in \D_{\RN{2}}\cap\{d\leq c\,\land\,
d=c\Rightarrow R_0< 1\}$ in the sense of Lemma \ref{Lem_SIRS}. Then there exists a unique dilatation transformation $(\xi,\la)\in\Gdil$ such that 
$\bx':=(\xi,\la)\re\bx\in\Dbio$ and such that the rescaled triangle satisfies $\T_{(\xi,\la)}=\Tph(B'_+,0)$. 
\item[ii)]
$\Gdil\re\Dsirs=\D_{\RN{2}}\cap\{d\leq c\,\land \,
d=c\Rightarrow R_0< 1\}$.
\end{itemize}
\end{proposition}

\begin{proof}
To prove part i) denote transformed quantities by a prime. The  requirements 
$\bT_\triangleleft'=(0,0)$ and 
$\bT_\vartriangle'=(0,1)$ fix 
$\xi=(1-X_\triangleleft)\inv$ and $\la=I_\vartriangle\inv$. Hence 
$X_\triangleright$ maps to 
$\xi cI_\vartriangle=c'=B'_+$ and therefore 
$\T_{(\xi,\la)}=\Tph(B'_+,0)$. To show $0\leq R'_i\leq B'_+$ use $R'_0=\xi(R_0-1)+1=\xi(R_0-X_\triangleleft)$ and therefore 
$$
0\leq R'_0\leq\frac{\xi}{\la}\min\{c,(c-d)/a\}=
\min\{c',(c'-d')/a'\}\leq B'_+
$$
By the above we also have 
$R'_1=a'R'_0+d'\leq c'=B_+'$ and we are left to show 
$R'_1\geq 0$. Sufficient is $d'\geq 0$ which follows from 
$1-d'/c'=\xi(1-d/c)\leq\xi(1-X_\triangleleft)=1$. This proves 
part i) and therefore also the 
``$\supset$''-direction of part ii). To prove the ``$\subset$''-direction use that 
%in $\Dsirs$ we have $0\leq R_i\leq B_+=c$ and hence $d-c=R_1-aR_0-B_+\leq 0$, where equality implies $R_0=0$ and $R_1=B_+$.
the action of $\Gdil$ on $\D$ preserves the sign of $d-c$ and in case $d=c$ we have $R_0=0$ and therefore
$R'_0=\xi(R_0-1)+1=1-\xi<1$. 
\end{proof}
As in \autocite{Nill1}, the above dilatation symmetry leads to a second normalization step for the SIRS-Sector, thus further reducing its number of essential parameters from four to two.
Equivalently this means, that equivalence classes of $\Gdil$-isomorphic systems with parameters in $\Gdil\re\Dsirs$ are naturally parametrized by 
$\Ksirs:=(\Gdil\re\Dsirs)/\Gdil$. A convenient realization of the  normalized system on phase space
$\P=\{(q,p)\in\RR\times\RRN\}$ is given by putting
\begin{equation}
q(t):=\frac{1}{a}( X(t/a)-1)\,,\qquad
p(t):=\frac{c}{a}I(t/a)
\label{uv}
\end{equation}
In terms of these variables the RN-dynamical system 
\eqref{dot_X} - \eqref{dot_Y} becomes
\begin{equation}
\dot{q}	=-q(p+1)+\kappa_0-\kappa_1 p\,,\qquad
\dot{p}	= qp\,,
\label{dot_qp}
\end{equation}
where the new $\Gdil$-invariant parameters are given by 
\begin{equation}
\kappa_0:=\frac{R_0-1}{a}\,,\qquad
\kappa_1:=\frac{c-d}{ac}\,.
\end{equation}
The only remaining constraint on the reduced parameter space says
\begin{equation}
\Ksirs=\{(\ka_0,\ka_1)\in\RR\times\RRN\mid\ka_1=0\Rightarrow\ka_0<0\}\,.
\end{equation}
Thus, after normalization the whole SIRS Sector just looks like Hethcote's classic endemic model except for a somewhat less restricted parameter space. In fact, by \Eqref{Heth2}, 
$\Dheth\subset\Dsirs$ is already two-dimensional with independent parameters $a\in(0,1)$ and $c=R_0=B_+>0$. These map injectively to 
$\Ksirs$ via
$\ka_0=(c-1)/a$ and $\ka_1=1/a$, whence
\begin{equation}
\Dheth\cong\Kheth=\Ksirs\cap\{\ka_1>1\,\land\,\ka_0+\ka_1>0\}
\end{equation}
The normalization convention in \Eqref{class_end} is obtained under the restriction $c>d$ or equivalently 
$\kappa_1>0$. In this case one may alternatively use
\begin{align}
u(t)-1	&:=\frac{c}{c-d}(X(ct/(c-d))-1)=\frac{1}{\ka_1} q(t/\ka_1)\,,\\
v(t)	&:=\frac{c^2}{c-d}I(ct/(c-d))=\frac{1}{\ka_1} p(t/\ka_1)\,.
\end{align}
In terms of these variables we recover the normalization convention \eqref{normalization}, \eqref{class_end}
\begin{equation}
\dot{u}=-uv-c_1u+c_2\,,\qquad\dot{v}=uv-v\,,
\end{equation}
where  $c_1=1/\ka_1$ and 
$c_2=1/\ka_1+\ka_0/\ka_1^2$, which is also the version given in \autocite{Nill1}. In part III of this work the above normalization step will be generalized to all Sectors of $\Dbio$. In this way the equation for $\dot{q}$ in \eqref{dot_qp} gets an additional term $-\kappa_2p^2$, and so our initial 14-parameter\footnote{i.e. constraint by $\nu_1=\nu_2$.} SSISS model boils down to a much simpler 3-parameter dynamical system.

\section{The case $\al_1=\al_2=0$\label{Sec_a=0} }

This Appendix shortly discusses the border case $\al_1=\al_2=0$\footnote{Here, for simplicity of notation, the tilde is still omitted. So beware that truly this appendix addresses the cases 
$\al_i=\nu_i=\nu_I=0$ (constant population \eqref{notation}) or 
$\al_i=\delta_i=0$ and $\mu_1=\mu_2$ (time varying population \eqref{notation2}).}.
In this case define parameter spaces $\C_x^0$ as in Eqs. \eqref{C}-\eqref{Cbio} with $\al_i=0$ and 
$\A_x^0:=\C_x^0\times\B$. In particular, in $\Abio^0$ we have 
$\theta_1\geq 0$, $\theta_2=0$, $\ga_i\geq 0$ and $\ga_1+\ga_2=1$. Lemma \ref{Lem_Xrep-ODE} 
still holds with $a=b=0$ and $d=R_1+\ep$, i.e. the replacement number dynamics becomes 
\begin{equation}
\dot{X}=(d-cX)I-\ep I^2\,,\qquad
\dot{I}=(X-1)I\,.
\label{dyn_a=0}
\end{equation}
In this case $R_0$ is undefined and there is a continuum of disease free equilibria at $I=0$, which are locally stable for $X<1$ and unstable for $X>1$. Proposition \ref{Prop_SS'} remains unchanged provided $\bfa,\bfa'\in\A^0$. Putting 
$\D^0=\{(c,d,\ep)\in\RR^3\}$ Lemma \ref{Lem_new-par} still holds with $\A$ replaced by $\A^0$ and $\D$ replaced by $\D^0$.
Moreover, in $\Abio^0$ we get $A_+=B_+=\be_1+\theta_1$, $A_-=B_-=\be_2$, $c=\be_1+\be_2+\theta_1$, $\ep=\be_2(\be_1+\theta_1)$ and 
putting $\D_A^0=\D_B^0=\DAB^0:=\D^0\cap\{c^2>4\ep\}$  Proposition \ref{Prop_Aph+Abio} becomes
\begin{align}
\phi(\Aph^0)	&=(\D_B^0\times\B)\cap
\{\be_2\leq \{B_-,R_1\}\leq\be_1\leq B_+\}\,,
\label{A0_ph}\\
\Dph^0	&=\D_B^0\cap\{R_1\leq B_+\}\,,
\label{D0_ph}\\
\phi(\Abio^0)		&=(\D_B^0\times\B)\cap
\{B_-=\be_2\leq R_1\leq\be_1\leq B_+\}\,,
\label{A0_bio}\\
\Dbio^0	&=D_B^0\cap\{B_-\leq R_1\leq B_+\}\subset\Dph^0\,.
\label{D0_bio}
\end{align}
So, for $\bx\in\Dph^0$ physical triangles $\Tph(\be_1,\be_2)$ are forward invariant provided $(\be_1,\be_2)$ satisfy the bounds \ref{A0_ph}. Finally, \Eqref{i_B(D_B)} becomes 
$\phi\inv(i_B(\D_B^0))=\phi(A_{\bth=0}\cap\A_{\bal=0})$ and Theorem \ref{Thm_periodic}, Theorem \ref{Thm_gaugefix} and Corollary \ref{Cor_periodic} stay valid also for $\al=0$.

\bsn
{\bf Acknowledgement}
I would like to thank Florin Avram for encouraging interest and useful discussions.

%\newpage
\printbibliography

\end{document}